\documentclass[12pt,preprint]{aastex}
\def\gsim{{{}_>\atop{}^{{}^\sim}}} \def\lsim{{{}_<\atop{}^{{}^\sim}}}
\def\eq#1{equation~(\ref{#1})} \def\Eq#1{Eq.~\ref{#1}}
\def\fig#1{Figure~\ref{#1}} \def\keywords{} 
    
 \def\aap{A\&A} \def\apjl{ApJ} \def\apjs{ApJS}

\begin{document}

\def\newpage{\vfill\eject} 
\def\vs{\vskip 0.2truein}
\def\msun{M_\odot} 
\def\rsun{R_\odot} 
\def\met{[M/H]} 
\def\vi{(V-I)}
\def\mtot{M_{\rm tot}} 
\def\mhalo{M_{\rm halo}} 
\def\pp{\parshape 2 0.0truecm 16.25truecm 2truecm 14.25truecm}
\def\la{\mathrel{\mathpalette\fun <}}
\def\ga{\mathrel{\mathpalette\fun >}}
\def\fun#1#2{\lower3.6pt\vbox{\baselineskip0pt\lineskip.9pt
\ialign{$\mathsurround=0pt#1\hfil##\hfil$\crcr#2\crcr\sim\crcr}}}
\def\kpc{{\rm kpc}} 
\def\Mpc{{\rm Mpc}} 
\def\kmsec{{\rm km/sec}}
\def\ibl{{\cal I}(b,l)} 
\def\kms{{\rm km}\,{\rm s}^{-1}}
\def\dii{\delta I/I} 
\def\di0{(\delta I/I)_0} 
\def\tp{t_{\rm peak}}
\def\day{\rm day}
\def\days{\rm days} 
\def\tp{t_{\rm peak}}
\def\td{{t_{\rm day}}} 
\def\thetae{\theta_{\rm E}} 
\def\ftot{F_{\rm tot}} 
\def\nr{N_{\rm r}} 
\def\ci{{\cal I}} 
\def\taus{\tau_{\rm start}}
\def\taue{\tau_{\rm end}}
\def\ch{{\bf CHANGED }}

\lefthead{GAUDI \& LOEB} \righthead{RESOLVING GAMMA-RAY BURST
AFTERGLOWS}
%%%%%%%%%%%%%%%%%%%%%%%
%%%%%%%electronic submission format
%\submitted{Version of \today}
\title{Resolving the Image of Gamma-Ray Burst Afterglows with
Gravitational Microlensing} 
\author{B. Scott Gaudi\footnote{Hubble Fellow}} 
\affil{Institute for Advanced Study, Einstein Drive, Princeton, NJ 08540} 
\authoremail{gaudi@sns.ias.edu} 
\author{Abraham Loeb} 
\affil{Harvard-Smithsonian CfA, 60 Garden Street, Cambridge, MA 02138} 
\authoremail{aloeb@cfa.harvard.edu}

%%%%%%%%%%%%%%%%%%%%%%%%%%%%%%

\begin{abstract}
Microlensing of a gamma-ray burst afterglow by an intervening star can
be used to infer the radial structure of the afterglow image.  Near
the peak of the microlensing event, the outer edge of the image is
more highly magnified than its central region, whereas the situation
is reversed at later times due to the rapid radial expansion of the
image on the sky.  Thus, the microlensed afterglow light curve can be
inverted to recover the self-similar radial intensity profile of the
afterglow image.  We calculate the expected errors in the recovered
intensity profile as a function of the number of resolution elements,
under the assumption that the afterglow and microlensing event
parameters are known.  For a point-mass lens and uniform source, we
derive a simple scaling relation between these parameters and the
resultant errors. We find that the afterglow need not be monitored for
its entire duration; rather, observations from the peak magnification
time $\tp$ of the microlensing event until $\sim 7\tp$ are sufficient
to resolve the majority of the afterglow image.  Thus, microlensing
events can be alerted by relatively infrequent observations of
afterglows and then monitored intensively, without significant loss of
information about the afterglow intensity profile. 
The relative intensity profile of $\sim 1\%$ of all afterglows
can be measured with 10 resolution elements to an accuracy of
${\cal O}(1\%)$ in the optical and ${\cal O}(10\%)$ in the infrared,
using 4m-class telescopes.  Weak microlensing events with large impact
parameters are more common; we estimate that for $\sim 10\%$ of
afterglows the image profile may be inverted to a fractional
accuracy $\la 20\%$ through frequent optical observations.  We also
calculate the effects of external shear due to the host galaxy or a
binary companion, and contamination by background light from the host
galaxy. 

\end{abstract}

%%%%%%%%%%%%%%%%%%%%%%%%%%%%

\setcounter{footnote}{0}
\renewcommand{\thefootnote}{\arabic{footnote}}

\section{Introduction\label{sec:intro}}

Recent detections of afterglow emission in the X-ray, optical and radio has
revolutionized gamma-ray burst (GRB) astronomy (for a recent review, see
\citealt{kulk2000}).  The afterglow emission appears to be reasonably
well-described by the fireball model \citep{wax1997a}, in which a
relativistically expanding shell of hot gas encounters an external
quiescent medium.  A blast wave is created, sweeping up the ambient
material, and accelerating relativistic electrons which then emit
synchrotron radiation \citep{pr1993,mr1997}. In this model, the spectral
flux of the afterglow is a broken power-law function of frequency which
evolves with time in a way that depends on the physical parameters of the
fireball and its surrounding medium, such as the total energy output of the
GRB source, the collimation angle of the outflow, the ambient gas density,
the fraction of the post shock energy which is converted into magnetic
fields and accelerated electrons, and the energy distribution of the
accelerated electrons \citep{spn1999}.  So far, observations of the
afterglow spectrum over a wide range of frequencies and times was the
primary method used to place constraints on these parameters (e.g.\
\citealt{wg1999,pandk2001,fandw1999}).  Crude constraints on the size of
the emitting region at late times were derived based on considerations
involving synchrotron self-absorption \citep{kandp1997} and radio
scintillations due to the intervening interstellar medium of the Milky-Way
galaxy \citep{g1997,wkf1998}.

The afterglow image is predicted to appear as a thin ring on the sky
\citep{wax1997c,sari1998,pm1998} at frequencies near or above the peak of
the synchrotron emission \citep{gps1999}, but more like a uniform disk
around or below the synchrotron self-absorption frequency \citep{gps1999b}.
In between breaks in the power-law spectrum of an afterglow, the contrast
and width of this ring evolve self-similarly, i.e.\ the brightness profile
maintains its shape as a function of radius when the latter is normalized
by the expanding circular boundary of the image \citep{gps1999}. The
self-similar brightness profile of the image changes only across spectral
breaks \citep{gl2001}.

The outer radius of the image expands radially at an apparent superluminal
speed of $\sim \Gamma c$, where $\Gamma$ is the Lorentz factor of the
emitting shock. The image occupies an angle of $1/\Gamma$ relative to the
center of the explosion due to relativistic beaming, and hence has a radius
$R_{\rm s}\propto r_{\rm sh}/\Gamma$, where $r_{\rm sh}$ is the shock
radius.  Geometric time delay implies that the observed time, $t \propto
r_{\rm sh}/\Gamma^2$, and so the radius scales as $R_{\rm s} \propto \Gamma t$.  For an ambient
density profile $\rho \propto r^{-k}$, one gets $\Gamma\propto r_{\rm
sh}^{-(3-k)/2}$. Therefore, $R_{\rm s}\propto t^{\delta}$ with
$\delta\equiv (5-k)/[2(4-k)]$.  For a uniform ambient medium with $k=0$
\citep{bm1976}, $R_{\rm s}\propto t^{5/8}$, while for a wind profile with
$k=2$ one gets $R_{\rm s}\propto t^{3/4}$.
The predicted differences between the afterglow images in these two cases
are typically small \citep{gl2001}.\footnote{At very late times, of order
months after the GRB trigger, the fireball becomes non-relativistic. In
this regime, the observer sees the entire fireball which follows the
Sedov-Taylor solution with $R_{\rm s}\propto t^{2/5}$.}

By resolving the image of a GRB afterglow at different frequencies,
one can obtain precious, new constraints on the fireball model.  As
originally pointed out by \citet{landp1998}, after about one day, the
angular size of a GRB ring is typically a micro-arcsecond ($\mu$as),
i.e. of the same order as the angular Einstein ring radius of a solar
mass lens at cosmological distances,
\begin{equation}
\thetae=\left( {4GM \over c^2D}\right)^{1/2} = 1.6 \left({M \over
M_\odot}\right)^{1/2}\left({D \over 10^{28}~{\rm cm}}\right)^{-1/2}
\mu{\rm as},
\label{eqn:thetae}
\end{equation}
where $M$ is the lens mass, and $D\equiv D_{os}D_{ol}/D_{ls}$, and
$D_{os}$, $D_{ol}$ and $D_{ls}$ are the angular diameter distances between
the observer-source, observer-lens, and lens-source, respectively.  This
fortuitous coincidence, along with the fact that the afterglow image
expands superluminally, means that lensing by an intervening solar-mass
star will produce a detectable deviation in the afterglow light curve with
a duration suitable for intense monitoring.  As the afterglow image
expands, different parts of it sweep by the lens and get amplified.  The
differential magnification of the source image can be used to recover the
radial structure of the afterglow image through a series of relative flux
measurements \citep{landp1998,mandl2001,gl2001}.  The probability that a
source at redshift $z\sim 2$ will have a projected angular distance less
than the Einstein radius of any intervening star (the optical depth) is
$\tau \sim 0.3 \Omega_*$, where $\Omega_*$ is the cosmological density of
stars in units of the critical density.  The value of $\Omega_*$ is
constrained to be $\ga 1\%$ based of the known populations of luminous
\citep{fhp1998} and faint \citep{alcock2000} stars and $\la 5\%$ based on
Big-Bang nucleosynthesis. Hence, $\tau \sim 1\%$
\citep{pandg1973,bandw1992,kandw2001}, and roughly one out of a hundred
afterglows should be strongly magnified (although all events may show
weak magnification signals due to a star located at ten Einstein radii from
their line-of-sight; see \citealt{mandl2001}).  

Intense monitoring of lensing events could provide invaluable information
about the nature of GRB afterglows.  Recently, \citet{gls2000} interpreted
the unusual optical afterglow light curve of GRB 000301C \citep{sagar2000,
randf2000} as being microlensed by an intervening $\sim 0.5M_\odot$ star.
They found that the optical-infrared light curve was well-fit by a model in
which the emission originated from a ring of fractional width $\sim 10\%$,
in agreement with earlier theoretical predictions.  \citet{pana2001} argued
that, if one adopts realistic surface brightness profiles for the afterglow
image, the optical/infrared afterglow of GRB 000301C cannot be explained by
microlensing.  \citet{ggl2001} showed, however, that microlensing can
reproduce the observed light curve, provided that the source is
significantly limb-brightened, namely that $\gsim 60\%$ of the flux arises
from the outer 25\% of the area of the afterglow.  This requirement is met
by the surface brightness profiles expected for emission frequencies above
the cooling break frequency \citep{gl2001}. Unfortunately, the light curve
coverage in radio frequencies was too sparse to uniquely confirm the
microlensing interpretation through, e.g.\ comparison of the inferred
surface brightness profiles across spectral breaks. 

The idea of using microlensing to study the images of gamma-ray burst
afterglows is similar in spirit to the idea of using microlensing to
resolve the atmospheres of stars in the Magellenic clouds and the Galactic
bulge \citep{ls1995,valls1995,sass1997,hsl2000}; for a recent review, see
\citet{gould2001}.  For background stars, the source radius stays constant
and the resolution of the source occurs because the caustic structure of
the lens sweeps over the face of the star due to the relative source-lens
motion.  However, for GRB afterglows the relative source-lens motion is
negligible, and the resolution occurs as the image of the source itself is
expanding and sweeping by the lens.  Microlensing has been used to measure
limb-darkening of the stars in the Small Magellenic Cloud and the Galactic
bulge \citep{albrow1999, afonso2000, albrow2000, albrow2001a} and also to
measure the spatial variations of spectral lines across the face of two
stars \citep{alcock1997,castro2001,albrow2001b,hey2001}.  \citet{gandg1999}
considered the signal-to-noise requirements for resolving stellar
atmospheres with microlensing. Our goal here is to apply similar
considerations to GRB afterglows.

In this paper, we examine the observational constraints that can be
obtained on afterglow images through a concerted monitoring effort of the
GRB community in future microlensing events.  Specifically, we determine
the expected statistical errors in the recovered intensity profile as a
function of various parameters and input assumptions, assuming that the
afterglow and microlensing event parameters are known.  This significance
of this work is twofold: first, we show that the radial intensity profile
of afterglows can be determined with fairly good accuracy ($\lsim 10\%$)
for reasonable expense of observational resources, and second, we present
the effects of various assumptions on the error in the recovered intensity
profile, in order to test the robustness of the conclusions and to provide
guidelines for observers.  In \S\ \ref{sec:grbs} we present the formalism
for microlensing of GRB afterglows.  We describe the method for determining
the expected errors in the recovered intensity profile in \S\
\ref{sec:errors}, and apply it to a fiducial case in \S\ \ref{sec:scaling}.
In \S\ \ref{sec:inputs}, we determine the effects of various input
assumptions on the errors.  Finally, we summarize the implications of our
primary results and discuss the effects of some of our (unavoidably)
simplified assumptions in \S\ \ref{sec:conclusions}.

For concreteness, we will primarily focus on the scaling laws of a
spherical fireball in a uniform medium, but derive a general scaling
relation for the recovered errors in the relative intensity profile that is
applicable for any power-law external density profile, $\rho \propto
r^{-k}$.  The two currently popular models, $k=0$ (uniform medium) and
$k=2$ (stellar wind) yield qualitatively similar conclusions.  We do not
consider jetted outflows, although a collimated outflow of opening angle
$\theta_{\rm jet}$ would behave as if it is part of a spherical fireball at
early time as long as $\Gamma(t)>1/\theta_{\rm jet}$ \citep{rhoads1997}.  We
also assume that the afterglow flux exhibits a single power-law slope,
which is appropriate if the observed frequency does not cross one of the
several breaks in the afterglow spectrum (see \citealt{gl2001}).  Note
that we do not assume any specific, model-dependent form for the intensity
profile, i.e. we do not evaluate the errors on the coefficients of some
parametric form for the profile.  Rather, we evaluate the minimum
attainable errors via direct inversion of the afterglow light curve.

\section{Gamma-Ray Burst Afterglows and Microlensing\label{sec:grbs}}

For a spherical fireball, the observer sees a circular source image
with an outer radius of angular size $\theta$ that evolves as,
\begin{equation}
\theta_{\rm s}(t)=\theta_{0} \td^{\delta},
\label{eqn:thgt}
\end{equation}
where $\theta_0$ is the angular size of the afterglow after $1~\day$, $\td$
is the observed time in days, and $\delta=[5-k]/[2(4-k)]$ for an external
medium with density profile $\rho \propto r^{-k}$.  The angular size
$\theta_{0}$ depends on the energy of the burst, the density of the ambient
medium, and the redshift of the burst.  For typical parameters, $\theta_0$
is of order $1\mu{\rm as}$.  As long as an observed frequency, $\nu$, does
not cross any of the time-dependent spectral break frequencies, the
afterglow flux at that frequency evolves as a power-law of time (e.g., Sari
et al. 1999)
\begin{equation}
F_\nu(t)=F_{0,\nu}\td^{-\alpha},
\label{eqn:foft}
\end{equation}
where $F_{0,\nu}$ is the flux after $1~\day$.  \citet{gl2001}
provide a comprehensive study of the predicted image profile in the
different spectral regimes.

Now assume that an intervening compact object lies at an angular
distance $\theta_b$ from the line-of-sight to the center of the
afterglow.  We may then normalize the angular radius of the afterglow
image in units of the Einstein radius of this object
(Eq. \ref{eqn:thetae}),
\begin{equation}
R_{\rm s}={R_0 \td^{\delta}},~\label{eqn:rhot}
\end{equation}
where
\begin{equation}
R_0 \equiv {\theta_{0} \over \thetae}.
\label{eqn:rho0}
\end{equation}
Since both $\theta_0$ and $\thetae$ are ${\cal O }(1 \mu{\rm as})$ for
typical parameters, $R_0$ is of order unity. The value of $R_0$ depends
very weakly on the GRB energy output and the ambient density normalization,
but is more sensitive to the lens mass and the source and lens redshifts
(for $k=0$, see Eqs. 2 \& 3 in Garnavich et al.\ 2000).

The magnification of an extended source is given by,
\begin{equation}
\mu(t)= {{\int d^2r \mu_0({\bf r})I_\nu(t;{\bf r})} \over{{\int d^2r
I_\nu(t;{\bf r})}}},\label{eqn:aes0}
\end{equation}
where $\mu_0({\bf r})=(r^2+2)/(r\sqrt{r^2+4})$ is the point-source
magnification at vector position ${\bf r}$ relative to the lens, and
the integral is over the area of the source.  For lensing of a uniform
circular source of radius $R_{\rm s}$ by a single point mass with no
external shear, the magnification obtains the value $\mu(t; R_{\rm
s},b)=\Psi[r=R_{\rm s}(t),b]$, where \citep{bible, wandm1994},
\begin{equation}
\Psi(r,b)={2 \over \pi r^2} \left[ \int_{|b-r|}^{b+r} dR
{{R^2+2}\over{\sqrt{R^2+4}}} {\rm arccos}{{b^2+R^2-r^2}\over{2Rb}}+
H(r-b){\pi\over2}(r-b)\sqrt{(r-b)^2+4}\right].
\label{eqn:aus}
\end{equation}
Here $H(x)$ is the step function and $b\equiv \theta_b/\theta_E$ is
the angular separation (impact parameter) between the source center
and the lens in units of the Einstein ring radius.

\begin{figure*}[t]   
\epsscale{1.0} \centerline{\plotone{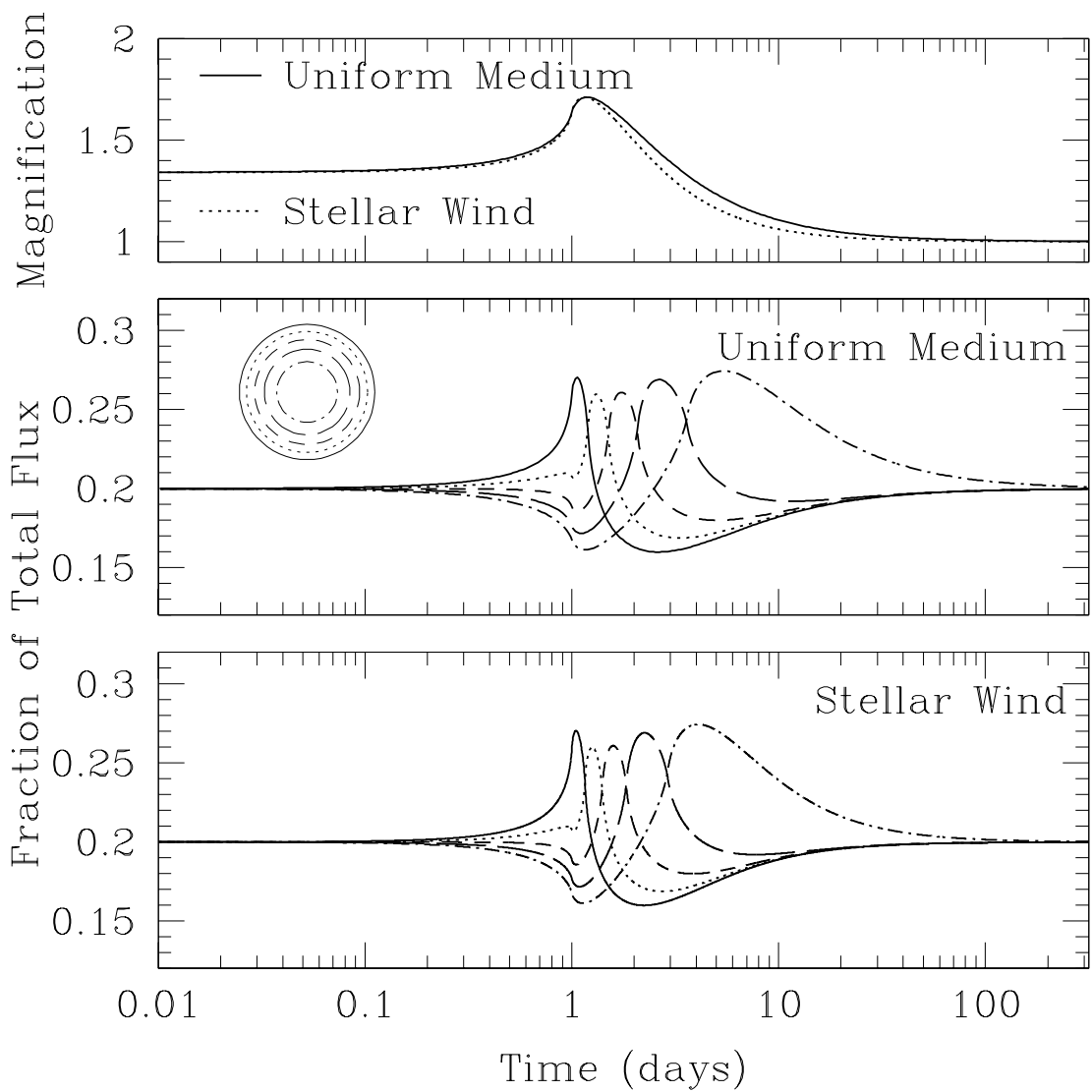}}
\caption{ \footnotesize The top panel shows the magnification as a function
of time for a uniform source with radius $R_0=1$ at $t=1~\day$ and an impact
parameter $b=1$, assuming a uniform external medium (solid line, $\rho={\rm
constant}$) and a medium such as that created by a stellar wind (dotted
line, $\rho\propto r^{-2}$).  The middle panel shows the fraction of total
flux contributed by five equal area annuli as a function of time in days
for the uniform external medium, while the lower panel shows the same but
for the stellar wind medium.  }
\label{fig:fig2}
\end{figure*}

For a uniform source (as applicable below the synchrotron self-absorption
frequency; see Granot et al.\ 1999b and Granot \& Loeb 2001), the
magnification history by a point-mass lens is completely specified by two
parameters: $b$ and $R_0$.  Figure \ref{fig:fig2} shows the magnification
as a function of time in days for $R_0=1$ and $b=1$, and two density
profiles, uniform ($\delta=5/8$) and $\rho\propto r^{-2}$ ($\delta=3/4$).
At early times when the source radius is small, the magnification is
roughly fixed at its point-source value
$\Psi_0(b)=\mu_0(b)=(b^2+2)/(b\sqrt{b^2+4})$.  The peak magnification
occurs at a time $\tp$ when $R(\tp)=b$, i.e. $\tp \approx
(b/R_0)^{1/\delta}~\days$.  The total flux from the afterglow is then
by,
\begin{equation}
F_{\rm tot,\nu}(t)=F_{0,\nu}\Psi(t;R,b)\td^{-\alpha}+F_{\rm
bg},
\label{eqn:ftot}
\end{equation}
where we have allowed for flux $F_{\rm bg}$ from any unresolved
sources not being lensed, e.g.\ the host galaxy of the GRB.  Thus, in
the simplest scenario the flux at a given frequency is a function of
five parameters: $F_{0,\nu},\alpha,F_{\rm bg},R_0$, and $b$.

In order to illustrate how microlensing effectively resolves the image of
the GRB afterglow, we plot in \fig{fig:fig2} the fraction of $\ftot$
contributed by five equal area annuli as a function of time assuming
$b=1,R_0=1$, and a uniform source.  For $t\lsim \tp$, the annuli contribute
roughly equal flux, and the source is not resolved.  However, beginning at
$t\sim \tp$ different annuli obtain different weights and the lens
differentially magnifies the source.  The annulus that contributes most of
the flux at a given time has a radius of $r_{\rm peak}=r_0 t^{\delta}\sim
b$ in units of $\thetae$, where $r_0$ is the radius of that annulus at
$t=1~\day$.  
Defining the fractional radius $X\equiv r/R_{\rm s} \equiv
r_0/R_0$,
\begin{equation}
X_{\rm peak}= \left( { t \over \tp}\right)^{-\delta}.
\label{eqn:xmax}
\end{equation} From inspection 
of \fig{fig:fig2} and \eq{eqn:xmax}, it is clear that
the the light curve need only be monitored from $\tp$ until
$0.3^{-1/\delta}\tp$ to resolve the outer $70\%$ of the radial profile
of the afterglow image.

\section{Error Analysis\label{sec:errors}}

Consider an azimuthally symmetric afterglow divided into $\nr$ annuli,
with annulus $i$ centered at radius $r_i$.  We assume a self-similar
behavior for which both the fractional radii $X_i$ and the mean
relative intensity $\ci(r_i)\equiv I(t;r_i)/I(t)$ in each annulus $i$
are constants in time.  The total flux from the afterglow is simply
the sum of the fluxes from each annulus.  The weight of each annulus
is in turn the area of the image of each annulus of width $\Delta
r_i$,
\begin{equation}
\Omega(t;r_i)=\pi[\Psi(t;r_i+\Delta r_i)(r_i+\Delta
r_i)^2-\Psi(t;r_i-\Delta r_i)(r_i-\Delta r_i)^2],
\label{eqn:areai}
\end{equation}
times the mean intensity of that annulus.  Converting \eq{eqn:aes0}
from an integral to a finite sum, the total flux is simply,
\begin{equation}
\ftot(t) = F_{\rm bg} + I(t) \sum_i^{\nr} \Omega (t;r_i) \ci(r_i).
\label{eqn:ftotf}
\end{equation}
The coefficients one wishes to determine are the $N_r$ relative intensities
$\ci(r_i)$.  These can be determined through inversion of the observed flux
$\ftot(t)$ if $F_{\rm bg}, I(t)$, and $\Omega(t;r_i)$ are known.  The
functions $I(t)$ and $\Omega(t;r_i)$ in turn depend on the parameters
$F_{0,\nu}, \alpha, b$, and $R_0$.  Schematically, the inversion can be
done as follows: $\alpha$ and $F_{\rm bg}$ can be determined from the
late-time ($t\gg \tp$) behavior of the afterglow light curve, when the
magnification due to microlensing is negligible.  The unlensed flux of the
afterglow could then be extrapolated back to $t=1~\day$, to determine
$F_{0,\nu}$.  The offset between the lensed and unlensed flux at early
times $t\ll \tp$ then gives $b$.  Finally, a measurement of $b$ combined
with $\tp$ provides $R_0$.  Therefore, all the parameters necessary to
invert \eq{eqn:ftotf} can ideally be determined without reference to the
surface brightness distribution of the source image.  In practice, of
course, a global fit to all $5+N_r$ parameters must be done simultaneously,
and this will introduce correlations between the parameters $\ci(r_i)$ and
$F_{0,\nu},\alpha,F_{\rm bg},R_0, b$, in turn inflating the errors on the
measured values of $\ci(r_i)$.  While $\ftot$ is linearly dependent on
$\ci(r_i)$ (see \Eq{eqn:ftotf}), this is not true for the parameters
$F_{0,\nu},\alpha,F_{\rm bg},R_0, b$.  The nonlinear dependences of these
parameters on the observed quantity $\ftot$ make the expected errors on
these parameters and their covariances with the parameters of interest,
$\ci(r_i)$, extremely difficult and time consuming to calculate.
Furthermore, the errors will depend sensitively on the exact coverage and
accuracy of the afterglow photometry.  For an observed microlensing event,
a full and careful determination of the errors on the inferred values of
$\ci(r_i)$ must account for their covariances with the other fit
parameters.  However, including these effects here is beyond the scope of
the paper.  Therefore, when computing the error bars on $\ci(r_i)$, we will
for simplicity assume that $F_{0,\nu},\alpha,F_{\rm bg},R_0, b$ are
perfectly known.  We stress that the errors we derive should thus be
regarded as lower limits to the actual errors.  

Now consider a series of flux measurements $\ftot(t_k)$ that are made at
times $t_k$ with errors $\sigma_k$, and that are fit to \eq{eqn:ftotf}.
The parameters of the fit are $\ci_i\equiv\ci(r_i)$, and the variances in
these parameters are $\delta\ci_i=({\cal C}_{ii})^{1/2}$, where ${\cal C}_{ij}$ is the
covariance matrix,
\begin{equation}
{\cal C}={\cal B}^{-1},~~~{\cal B}_{ij}=\sum_k \sigma_k^{-2} { {\partial
F(t_k)}\over{\partial \ci_i}} { {\partial F(t_k)}\over{\partial
\ci_j}},
\label{eqn:covar}
\end{equation}
and ${\partial F(t_k)}/{\partial \ci_i}=I(t_k) \Omega(t;r_i)$.  The
fractional errors are then $\delta\ci_i/\ci_i={\cal C}_{ii}^{1/2}/\ci_i$.  We
assume Poisson-noise limited precision, so that $\sigma_k \propto
\sqrt{F(t_k)}$.  Therefore, for a given set of assumptions about the
microlensing event ($b,R_0$), the unlensed flux of the afterglow
($F_{0,\nu}, F_{\rm bg}$) and the observational setup (duration of
observations, telescope diameter, etc.), the covariance matrix ${\cal C}$ can
be formed and the expected errors $\delta\ci_i/\ci_i$ determined.

\section{Results}

\subsection{Fiducial Scaling Relation\label{sec:scaling}}

We now apply the formalism in \S\ \ref{sec:errors} to derive an approximate
scaling relation for $\delta\ci_i/\ci_i$.  Rather than adopt specific (and
therefore model dependent) forms for the afterglow flux $F_{0,\nu}$ as a
function of frequency and a specific observational setup, we will simply
assume that the instrument being used collects $\Gamma_\nu$ photons per
second from the unlensed afterglow at $t=1~\day$.  In \S\
\ref{sec:discussion}, we will evaluate values of $\Gamma_\nu$ and thus the
errors expected from a specific example of afterglow emission and
observational setup.  For our fiducial scenario, we assume a uniform source
($\ci_i=1$), a negligible background flux ($F_{\rm bg}=0$), and that
observations are made continuously from the burst until very late times
$t\gg \tp$.  We also assume equal-area annuli, so that all bins have equal
weight for a uniform source.  We then calculate $\delta\ci_i/\ci_i$ for a
range of values of $\alpha$, $R_0$, $b$, and $N_r$.  Numerically, we find
the following approximate scaling relation for the fractional errors in the
recovered intensity profile,
\begin{equation}
\left({\delta\ci_i\over\ci_i}\right)_0= 0.35 \left({r_i \over
R_s}\right)^{ 5(1-\alpha)/8\delta + 3/2} \left({\Gamma_{\nu} \over
1~{\rm s}^{-1}}\right)^{-1/2} \left({N_r \over 10}\right)^{8/5}
R_0^{(1-\alpha)/2\delta} b^{\alpha/2\delta}.
\label{eqn:scale}
\end{equation}
Equation (\ref{eqn:scale}) is the primary result of this paper.  We find
that it predicts the expected errors to an accuracy of $\lsim 5\%$ for most
combinations of parameters.  Note that \eq{eqn:scale}, and indeed the
majority of the quantitative results in this paper, rely on four
assumptions about the afterglow, namely: (i) the unlensed flux of the
afterglow is a power-law function of time, $F\propto t^{-\alpha}$; (ii) the
radius of the afterglow image scales as $R_{\rm s}\propto t^{\delta}$;
(iii) the image profile evolves self-similarly; and (iv) the afterglow
image is circularly symmetric.  These assumptions should hold at least
approximately for some restricted part of most afterglow light curves.  We
discuss these assumptions more thoroughly in \S\ref{sec:conclusions}.  The
above equations can be used to determine the expected errors on the
recovered intensity from observations performed with various instruments,
apertures, and photon frequencies of afterglow light curves of arbitrary
fluxes and power-law indices.  Because we have assumed that the
observational precision is limited only by photon statistics, that there
are no correlations with the other parameters (see the discussion in \S\
\ref{sec:errors}), and that the measurements are continuous, equation
(\ref{eqn:scale}) provides the minimum attainable error, at least for
uniform sources and isolated lenses with no external shear.  In the next
section, we will relax some of the assumptions leading to this result in
order to evaluate their effect on the expected errors.  For the two cases
of a uniform medium ($\delta=5/8$) and a stellar wind medium
($\delta=3/4$), the exponents in \eq{eqn:scale} are quite similar.
Therefore, for the sake of simplicity, we will consider only the uniform
medium case with $\delta=5/8$, in the discussion that follows.

Several of the terms in \eq{eqn:scale} can be derived analytically, simply
by noting that, according to photon statistics, the errors should scale as
$\delta\ci_i/\ci_i \propto N_\gamma^{-1/2}$, where $N_\gamma$ is the number
of photons collected.  For example, the scaling with $R_0$ can be derived
as follows: the time of the peak of the light curve scales as $t_{\rm peak}
\propto R_0^{1/\delta}$.  The unlensed flux at this time is $F(t_{\rm
peak}) \propto R_0^{-\alpha/\delta}$.  The magnification structure is
independent of $R_0$; however, all times are scaled by $R_0^{1/\delta}$, so
the time over which photons can be collected scales as this factor.  Thus
the total error should scale as $(R_0^{1/\delta}
R_0^{-\alpha/\delta})^{1/2}=R_0^{(1-\alpha)/2\delta}$, as found
numerically.  The scaling with the impact parameter $b$ can be derived in a
similar fashion, under the assumption that the peak magnification scales as
$b^{-1}$. However, this is only strictly valid for $b\ll 1$.  Therefore,
the scaling $\delta\ci_i/\ci_i \propto b^{\alpha/2\delta}$ in
\eq{eqn:scale} is only approximate, and breaks down for $b\gg 1$.  Numerically, we find that for $b\le 4$,  \eq{eqn:scale} predicts errors that are too small by $\lsim 20\%$ for $r\gsim 0.5 R_s$ (the outer 70\% of the area of the image). 
The exponent for the scaling with $N_r$ is $8/5$, similar to that found by
\citet{gandg1999} for resolving the images of Galactic stars by
microlensing.  Naively, one might expect the prefactor in \eq{eqn:scale} to
be considerably smaller: for $\alpha=1$ and $b=1$ the total number of
photons collected is of order $N_\gamma\sim 10^{5} (\Gamma_\nu/1~{\rm
s}^{-1})$ (with only a logarithmic dependence on the integration time
during which the source is resolved).
Divided over $N_r=10$ bins, one might expect the error per
bin to be $(N_\gamma/N_r)^{-1/2} \sim 1\% 
(\Gamma_\nu/1~{\rm s^{-1}})^{-1/2}$, 
about an order of magnitude smaller than the $\sim 12\%(\Gamma_\nu/1~{\rm s^{-1}})^{-1/2}$ predicted by
\eq{eqn:scale} for a radius $r_i=0.5 R_s$.   
However, the errors increase with $N_r$ faster than one
would naively expect, and so the error in each bin is closer to
$(N_\gamma/N_r^{16/5})^{-1/2}\sim 12\% (\Gamma_\nu/1~{\rm s^{-1}})^{-1/2}$.

\subsection{Effects of Changes in the Input Assumptions\label{sec:inputs}}

\subsubsection{Duration of Observations\label{sec:dobs}}

\begin{figure*}[t]   
\epsscale{1.0} \centerline{\plotone{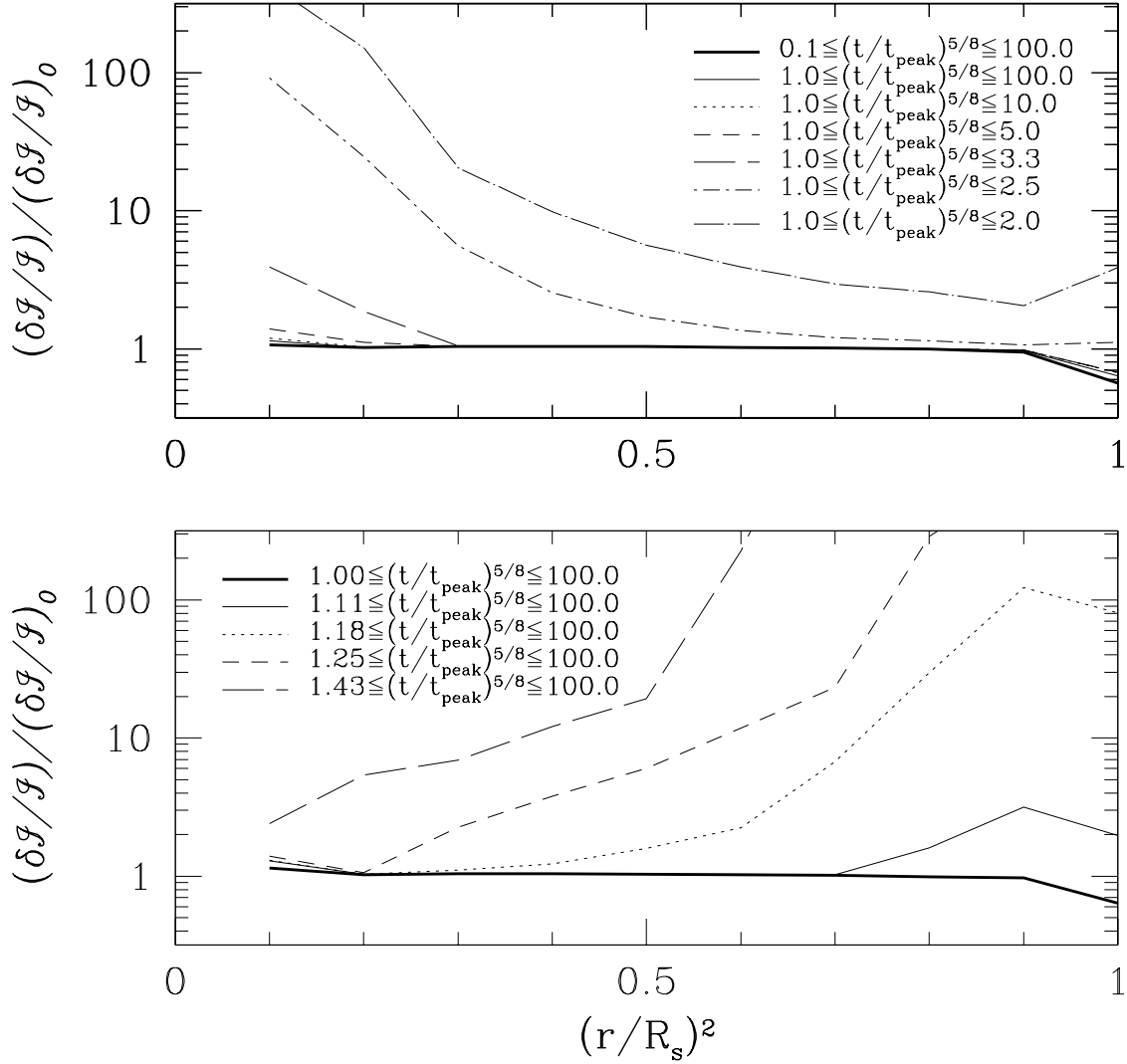}}
\caption{ \footnotesize Both panels show the fractional error in the
recovered intensity profile $\delta\ci/\ci$ normalized to 
the fiducial error $(\delta\ci/\ci)_0$ given by the scaling
relation in \eq{eqn:scale}, as a function of the square of the
normalized radius $X=r/R_{\rm s}$ of the afterglow.  Each line shows
various assumptions of the beginning and duration of
observations relative to the time $\tp$ of the peak of the
microlensing event.  {\it Top Panel}: Varying the end of observations
relative to $\tp$.  {\it Bottom Panel}: Varying the beginning the
observations relative to $\tp$.  }
\label{fig:fig3}
\end{figure*}

The most severe simplification made in deriving \eq{eqn:scale} is the
assumption of continuous measurement from the GRB trigger until long after
the peak time $\tp$.  Since the optical depth for microlensing is $\tau
\sim 1\%~b^2$, at least $100b^{-2}$ afterglows must be monitored to detect
one that has impact parameter $\le b$.  If it was truly necessary to
monitor all of these afterglows continuously to recover the intensity
profile accurately, this would represent an overwhelming observational
burden.  We therefore evaluate the errors expected under various
assumptions about the starting time and the duration of the observations.
We scale the starting time, $\taus$, and ending time, $\taue$, in terms of
the peak time through the relation $\tau\equiv(t/\tp)^{5/8}$, so that for
given $\taus$ and $\taue$, a fixed range of fractional radii, namely
$X_{\rm min}=\taue^{-1}$ through $X_{\rm max}=\taus^{-1}$, are probed
regardless of the values of $R_0$ and $b$ (see \Eq{eqn:xmax}).  The top
panel of \fig{fig:fig3} shows the expected errors $\delta\ci_i/\ci_i$
normalized to the fiducial errors in equation (\ref{eqn:scale}), for
$(\taus,\taue)=(0.1,100.0)$, corresponding to complete coverage of the
microlensing event, and also for $\taus=1$ (starting at the peak of the
event) and various values of $\taue$.  Note that here and throughout we
plot $\delta\ci_i/\ci_i$ as a function of $(r/R_s)^2$, because we assume
bins of equal area (and hence equal weight for a uniform source; see
Eq. \ref{eqn:areai}).  Clearly observations before the peak of the event
provide little information on the intensity profile.  Moreover, the
afterglow need only be monitored until $t\sim 7\tp$ to resolve the outer
two thirds of the source radius.  The bottom panel of \fig{fig:fig3} shows
the expected errors when $\taus$ is varied.  Delaying observations even
somewhat past the peak of the event will result in seriously degraded
information.

We conclude that measurements from $\tp$ until $\sim7\tp$ will provide
almost as much information as continuous measurements from the GRB trigger
until late times.  Since for typical parameters $\tp\sim 1~\day$, the
afterglow need only be aggressively monitored for about six days.  This
suggests the following detection strategy for maximizing the number of
lensed GRBs for which the intensity profile can be recovered.  Since the
sampling necessary to determine that the GRB is being microlensed is not as
dense as that needed to recover the intensity profile, a large number of
bursts can be monitored relatively infrequently from a global network of
small telescopes.  With real-time reduction, this network should be able to
``alert'' microlensing events before or near the peak time.  Larger
telescopes could then be used to densely sample the microlensed afterglow
for the $\sim 6$ days necessary to resolve the intensity profile.  This is
similar to the way Galactic microlensing observations are coordinated
\citep{udal1994,alcock1996}.

\subsubsection{Realistic Intensity Profiles\label{sec:iprofs}}

In this section, we explore the effects of realistic intensity profiles on
the resultant errors.  The optical image of the afterglow is expected to be
limb-brightened, because light from the edge of the afterglow suffers the
longest time delay, and thus was emitted at earlier times when the fireball
was brighter (Waxman 1997c; Sari 1998; Panaitescu \& Meszaros 1998; Granot,
Piran, \& Sari 1999a,b).  The contrast between the center and edge of the
afterglow, as well as the sharpness of the cut-off in intensity at the
outer edge of the afterglow, depends on the location of the observed
frequency relative to the various spectral breaks in the broken power-law
spectrum of afterglows (Granot et al.\ 1999a,b; Granot \& Loeb 2001).  The
exact shape of the relative intensity profile will affect the shape of the
light curve, as well as the resultant errors, with the errors expected to
be larger near the center (where the relative intensity is smaller) than
those calculated assuming a uniform source.  Previous studies of
microlensing \citep{landp1998,gls2000,mandl2001} have parameterized the
intensity profile as a uniform inner disk bounded by a uniform outer ring,
with two parameters specifying the width $W$ and contrast $C$ of the ring.
Of course, realistic intensity profiles do not exhibit such discontinuous
features.  The precise calculation of these profiles for the different
spectral regimes of GRB afterglows is presented elsewhere (Granot \& Loeb
2001).  Here, we adopt a parameterized form for the relative intensity
profile that captures the qualitative aspects of realistic profiles,
\begin{equation}
\ci(X) = \ci_0 \left[ 1 - c\sqrt{1-X^2}\right](1-X^n),
\label{eqn:iprof}
\end{equation}
where $c$ and $n$ are free parameters that define the shape of the
profile, and $\ci_0$ is the normalization such that $\int_0^1 XdX
\ci(X)=1$. The index $n$ defines the sharpness of the cut-off, and is
roughly analogous to the width of the ring $W$.  The coefficient $c$
is related to the intensity at $X=0$, and is roughly analogous to the
contrast $C$.  Note that for $c=0$ and $n\rightarrow \infty$, this
form reduces to a uniform source.

\begin{figure*}[t]   
\epsscale{1.0} \centerline{\plotone{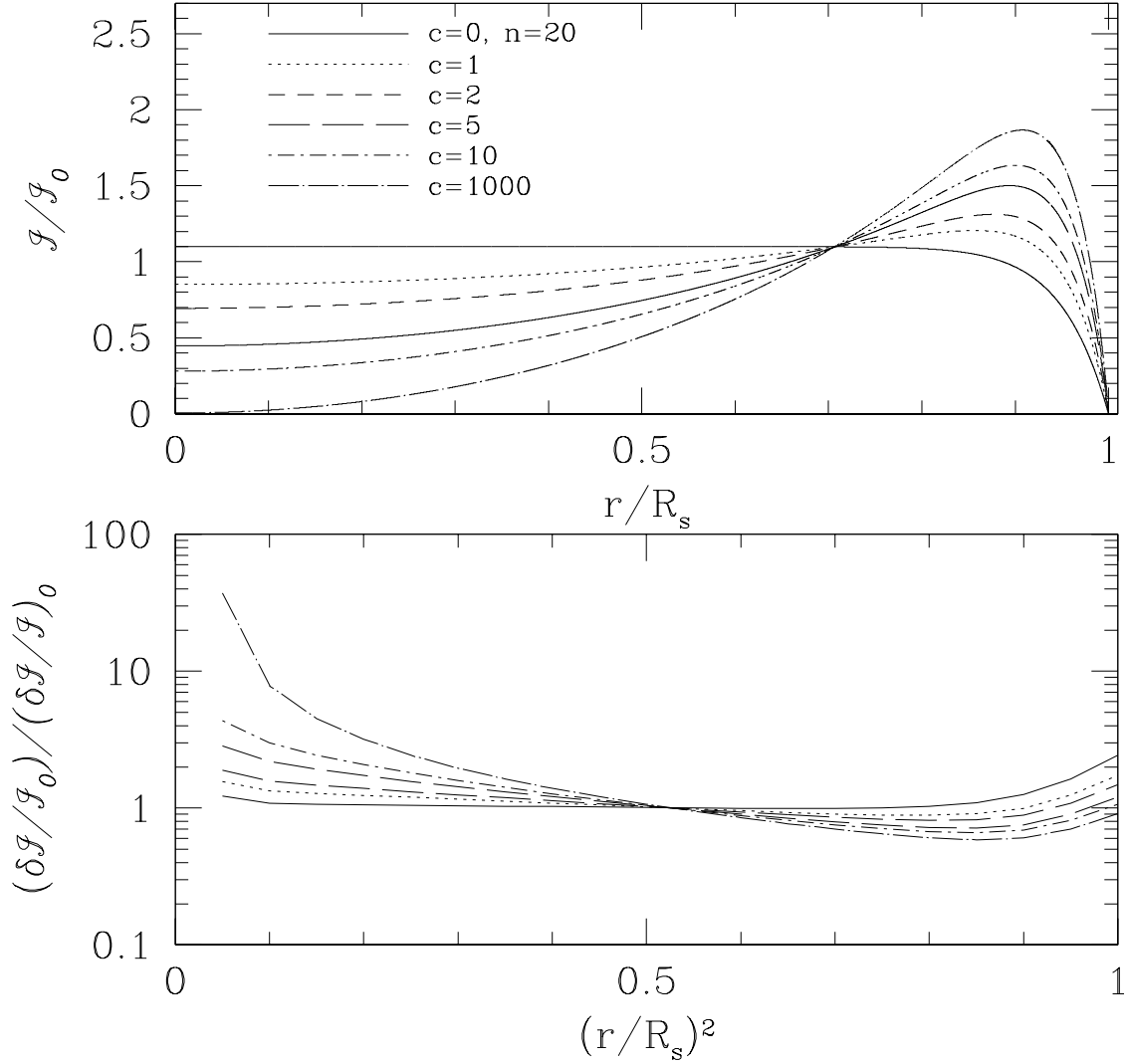}}
\caption{\footnotesize {\it Top Panel}: The normalized intensity
profile $\ci(r)/\ci_0$ of the source as a function of the normalized
radius $X=r/R_{\rm s}$, assuming a cutoff parameter $n=20$ and for
several different values of the parameter $c$ (see text and
\eq{eqn:iprof} for details).  {\it Bottom Panel}: The fractional error
in the recovered intensity profile $\delta\ci/\ci_0$ normalized to 
the fiducial error $(\delta\ci/\ci)_0$ given by the scaling
relation in \eq{eqn:scale}, as a function of the square of the
normalized radius of the afterglow, for the intensity profiles shown
in the top panel.}
\label{fig:fig4}
\end{figure*}

The top panel of \fig{fig:fig4} shows $\ci(X)$ for a relatively gradual
cutoff, $n=20$, and various values of $c$.  The bottom panel of
\fig{fig:fig4} shows the fractional errors $\delta\ci/\ci_0$ relative to
the fiducial errors for a uniform source in equation (\ref{eqn:scale}),
given the image profiles shown in the top panel.  Note that we normalize
the variances $\delta\ci$ to the mean intensity $\ci_0$ rather than the
intensity in that bin in order to avoid the fractional errors
$\delta\ci/\ci$ blowing up when $\ci \rightarrow 0$ for large $c$.  In
general, the resultant errors are a factor $\lsim 3$ worse than than the
uniform source case.  In the most extreme example where $\ci(X)\rightarrow
0$ at the center of the image, the fractional errors can become quite
large, but only for the innermost bin.  For the outer half of the area of
the image, the errors can be smaller than the uniform source case. Thus, in
this case, measurements will result in a robust upper limit to the
intensity at the center of the image, with a clear measurement for the
outer part of the image.  Therefore the ring-like structure of the image
will be accurately recovered.  We have also calculated, but do not show,
the errors for the case of a sharp outer cutoff, $n=500$.  The results for
this case are qualitatively and quantitatively similar to the results for
$n=20$. We conclude that realistic intensity profiles will not result in
significantly inflated errors relative to the uniform source estimate
(\Eq{eqn:scale}), at least for the majority of the source area.

\subsection{External Shear and Binary Lenses\label{sec:shear}}

We now consider more complicated models for the lens.  The majority of
lenses are likely to reside in galaxies, and a significant fraction of
these may be in binary systems.  Therefore, the assumption of an isolated
single lens with no external shear used in the previous sections will not
be valid.  We therefore consider two other types of lenses: a single lens
with external shear \citep{cha79}, and a binary lens \citep{sandw1986}.
The former case is appropriate for a lens perturbed by either a wide binary
companion or the potential of its host galaxy.  The latter is appropriate
for binary-lenses with separations ${\cal O}(\thetae)$ or smaller.  For a
lens residing in a region of high optical depth, i.e. near the center of a
galaxy, the magnification structure will not be well approximated as either
an isolated lens with external shear or a binary lens.  In this case one
may only be able to draw statistical inferences about the surface
brightness profile of the source, as it will be difficult to reconstruct
the magnification structure of the lens from the light curve alone.  We
will not consider this regime here. We compute the fractional errors in
the recovered intensity profiles in the same manner as before, except that
we must replace $\Psi$ in \eq{eqn:areai} with the uniform-source
magnification appropriate for the given lens model.  We compute this
magnification using the inverse ray-shooting method (see, e.g.,
\citealt{wambs1997}) as follows.  The generalized lens equation for point
masses and external shear can be expressed in complex coordinates on the
plane of the sky as \citep{witt1990},
\begin{equation}
\zeta=\gamma {\bar z} + z + \sum_k^N
{{\epsilon_k}\over{\bar{z}_k-\bar{z}}},
\label{eqn:lenseq}
\end{equation}
where $\zeta$ is the complex source position, $z$ is the complex image
position, $\gamma$ is the shear, $\epsilon_k$ is the fractional mass of
lens component $k$, and $z_k$ is the position of this component. All
distances are expressed in units of the angular Einstein ring of the
combined mass of the system.  For a single lens, $N=1$, $\epsilon_1=1$, and
$z_1=0$.  For a binary-lens with no external shear, $\gamma=0$, $N=2$, the
mass ratio is $q=\epsilon_1/\epsilon_2$, 
the dimensionless projected
separation is $d=|z_1-z_2|$, and we choose the origin to be 
the midpoint of
the binary. 

We sample the image plane uniformly and densely, solving for the source
position using the appropriate form of \eq{eqn:lenseq}.  These positions
are then binned in the source plane. The magnification of each bin is then
just the surface density of rays in the image plane divided by the surface
density of rays in the source plane.  
Since this procedure conserves flux, the resulting magnification map can
then be convolved with a source of arbitrary size to compute the
magnification of the extended source.  We calculate magnification maps for
$\gamma=0.1,0.2,0.3$ and $0.4$, and one binary-lens configuration with
$d=0.8$ and $q=1$.  For the single lens with external shear, $N_r+7$
parameters are needed to specify the light curve, namely: the $\nr$
relative intensities $\ci_i$, plus $F_{0,\nu},\alpha, F_{\rm bg},R_0$, the
shear $\gamma$, and two parameters that specify the position of the GRB
with respect to the lens, which we denote as $(x,y)$.  For the binary lens,
we assume that $\gamma=0$; however two additional parameters are needed to
describe the topology of the lens, namely $d$ and $q$, and so there are a
total of $N_r+8$ parameters.  When calculating the errors on the parameters
$\ci_i$, we will again assume that all other parameters are perfectly
known.  Here this assumption is somewhat less justified than the single
lens case, not only because there are more parameters, but also because the
effects of these parameters are in general not directly related to features
in the light curve, and are likely to be subject to degeneracies or
correlations, which will inflate the errors in the recovered $\ci_i$.  Such
denegeracies or correlations are known to inflate the errors in
limb-darkening measurements of stars using Galactic binary-lens events,
although not greatly so for well-measured light curves
\citep{afonso2000,albrow2001a}.  In the Galactic binary-lens case,
determination of all parameter combinations that correspond to solutions to
the observed light curve, and therefore identification of degeneracies and
correlations between fit parameters, is a formidable task
\citep{mdi1995,dip1997,albrow1999b}.  This is likely to also be true for
GRB afterglows.  A comprehensive study of parameter space is beyond the
scope of this paper, and so we will simply assume that the binary lens
parameters are known.  This assumption facilitates comparison with the
isolated single-lens case, which is the main purpose of our discussion
here.  We stress that the errors we derive under this assumption are almost
certainly underestimates.  In both the external shear and binary-lens case,
we will assume a uniform afterglow image (see Mao \& Loeb 2000 for other
examples).

\begin{figure*}[t]   
\epsscale{1.0} \centerline{\plotone{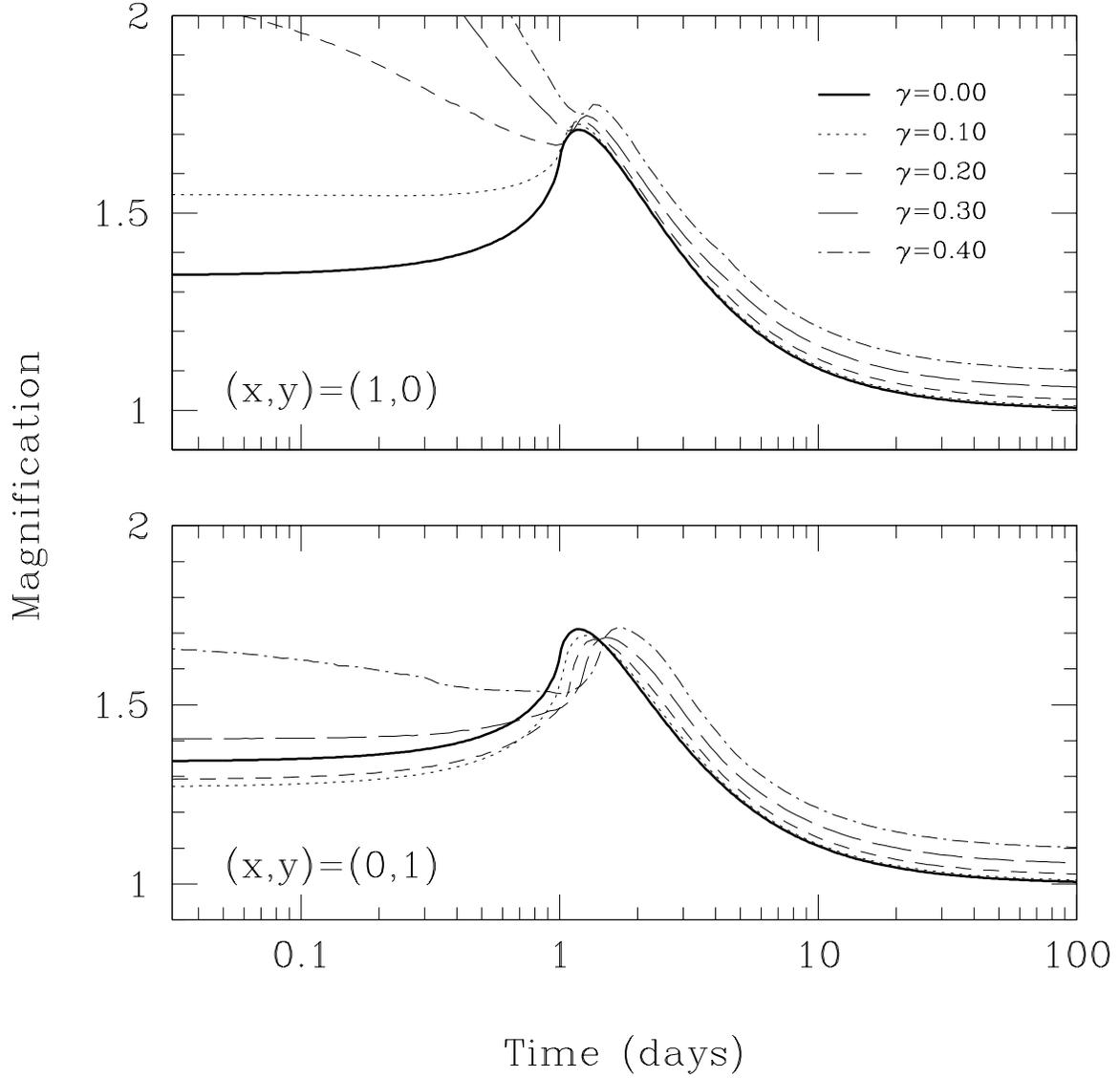}}
\caption{ \footnotesize Magnification as a function of time for a
uniform source with $R_0=1$ and $b=1$ for a single lens with external
shear along with $x$-axis of magnitude $\gamma$ for various values of
$\gamma$.  {\it Top Panel}: Assuming the center of the source is
located on the x-axis of the lens.  {\it Bottom Panel}: Assuming the
center of the source is located on the y-axis of the lens.}
\label{fig:fig6}
\end{figure*}

\begin{figure*}[t]   
\epsscale{1.0} \centerline{\plotone{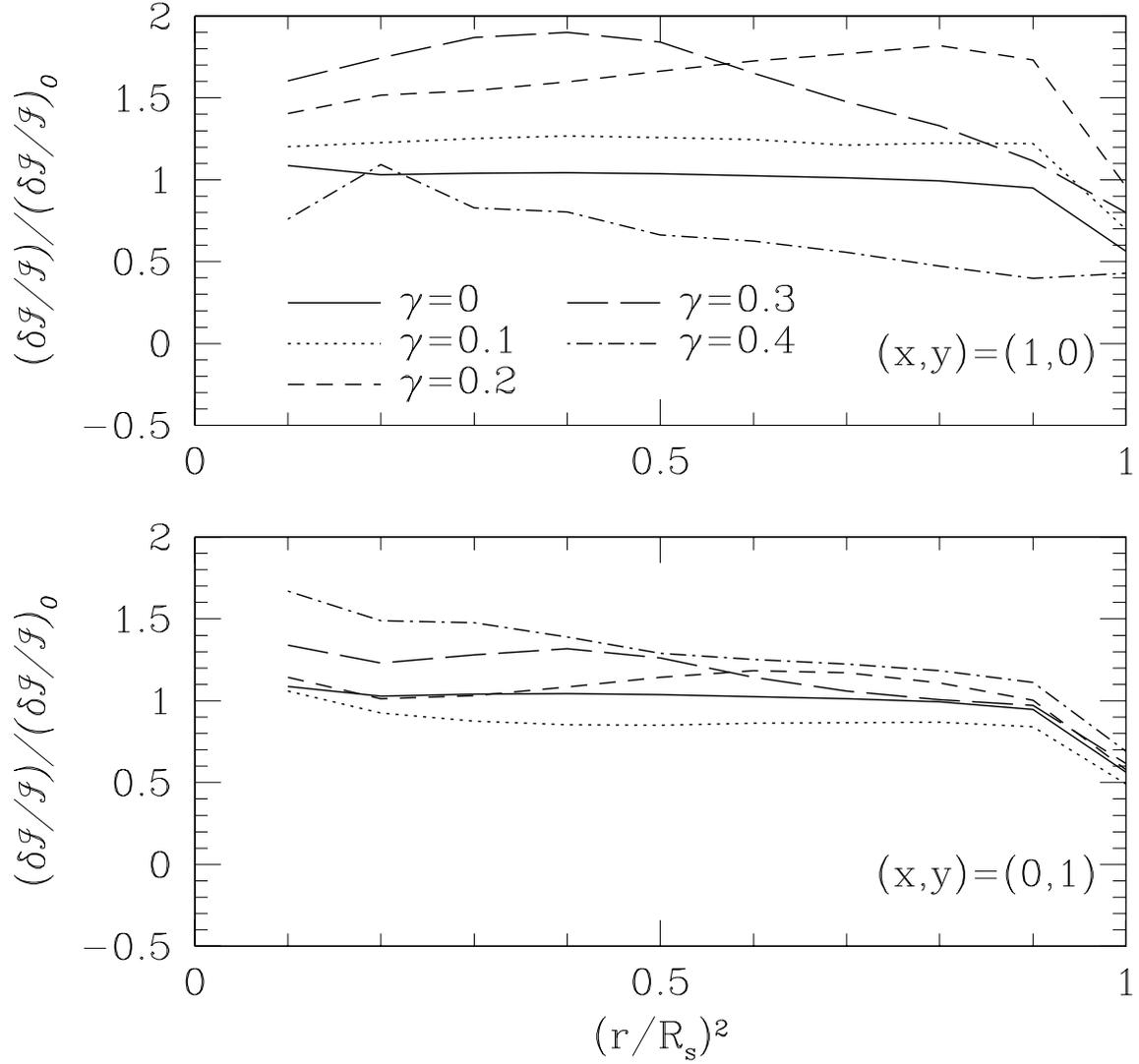}}
\caption{ \footnotesize The fractional error in the recovered intensity
profile $\delta\ci/\ci$ normalized to the fiducial error
$(\delta\ci/\ci)_0$ given by the scaling relation in \eq{eqn:scale}, as a
function of the square of the normalized radius $X=r/R_{\rm s}$ of the
afterglow for various values of the external shear $\gamma$ along the
x-axis.  We have assumed $R_0=1$, $b=1$, and a uniform source.  {\it Top
Panel}: Assuming the center of the source is located on the x-axis of the
lens.  {\it Bottom Panel}: Assuming the center of the source is located on
the y-axis of the lens.  }
\label{fig:fig8}
\end{figure*}

In \fig{fig:fig6} we show the magnification light curves for the case of a
single lens with external shear along the $x$-axis of magnitudes
$\gamma=0,0.1,0.2,0.3$, and $0.4$, both for a source centered on the x-axis
of the lens, $(x,y)=(1,0)$, and a source centered on the y-axis of the lens
$(x,y)=(0,1)$.  Note that the caustics are most highly elongated along the
x-axis, and thus for a given $\gamma$, light curves where the source is
centered on the x-axis should show more dramatic deviations from the
$\gamma=0$ case than for sources centered on the y-axis (Mao \& Loeb 2000).
\fig{fig:fig8} shows the resultant errors relative to the fiducial error
estimate in equation(\ref{eqn:scale}).  In general, sources centered on the
x-axis show more dramatic variations in the expected errors than sources
centered on the y-axis.  In all cases the fractional errors are within a
factor of $2$ of the $\gamma=0$ expectation.  Thus, at least for modest
values of $\gamma \le 0.4$, we conclude that an external shear will not
result in significantly inflated statistical errors, assuming that the
covariances with the afterglow and microlensing parameters are small. 

\begin{figure*}[t]   
\epsscale{1.0} \centerline{\plotone{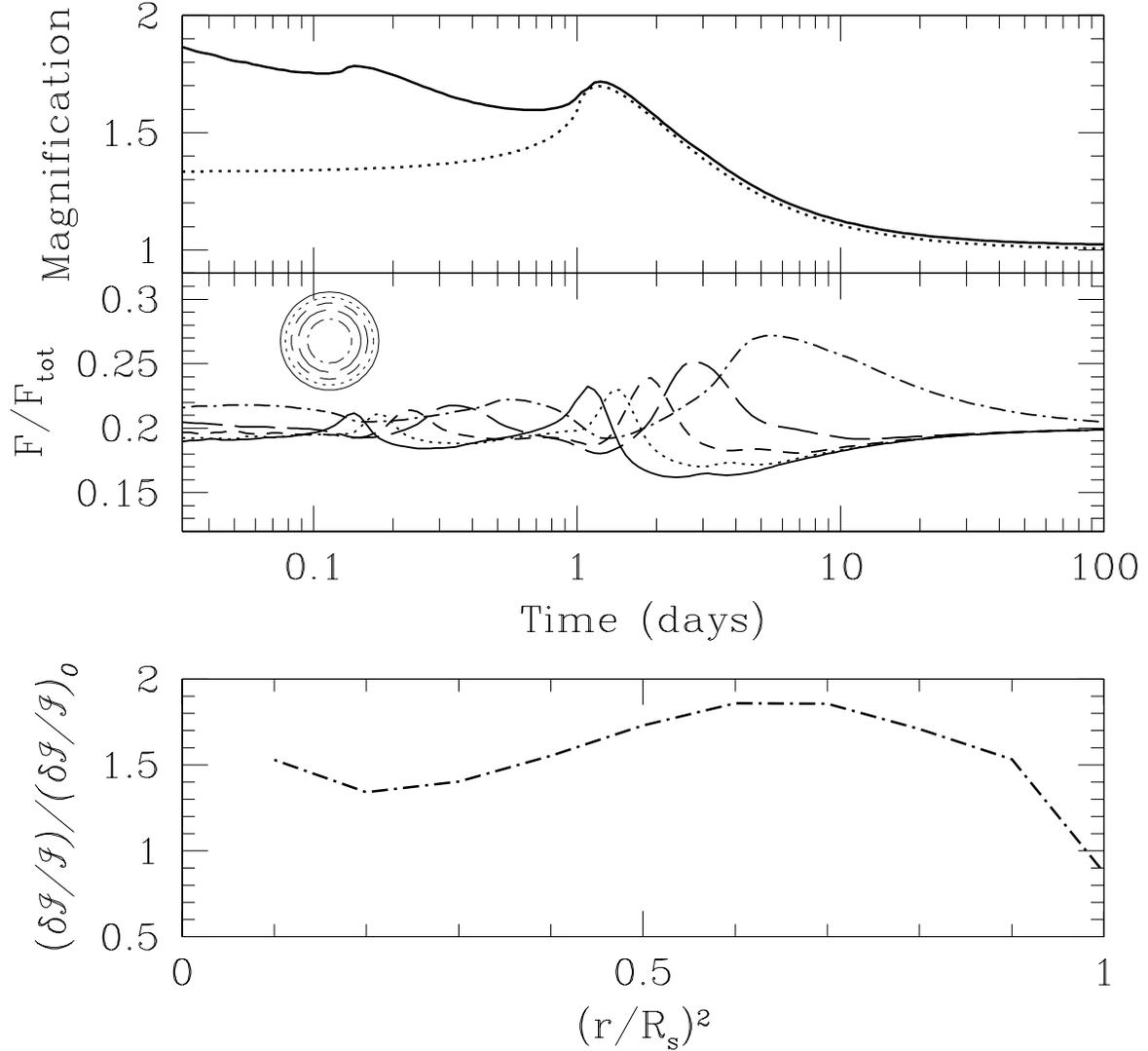}}
\caption{ \footnotesize {\it Top Panel}: The magnification as a function of
time for a uniform source with $R_0=1$.  The solid line is for a
binary-lens with mass ratio $q=1$ and a separation of $d=0.8$ in units of
the combined Einstein ring radius.  The source is centered on $(-0.16,-1)$.
The dotted line is for a single lens with impact parameter $b=(0.16^2 +
1^2)^{1/2}=1.013$.  {\it Middle Panel}: The fraction of total flux
contributed by five equal area annuli as a function of time in days for the
binary-lens light curve shown in the top panel.  {\it Bottom Panel}: The
fractional error in the recovered intensity profile $\delta\ci/\ci$
normalized to the fiducial error $(\delta\ci/\ci)_0$ given by
the scaling relation in \eq{eqn:scale}, as a function of the square of the
normalized radius $X=r/R_{\rm s}$ of the afterglow.}
\label{fig:fig9}
\end{figure*}

In \fig{fig:fig9}, we show the magnification as a function of time for
the binary lens with $d=0.8$ and $q=1$, and a source centered at
$(x,y)=(-0.16,-1)$.  We also show the fraction of total flux
contributed by five equal area bins as a function of time.  For times
$t>1~\day$, the light curves resemble that for a single lens (compare
with \fig{fig:fig2}), as the annuli become comparable to or larger
than the caustic features.  The bottom panel of \fig{fig:fig9} shows
the resultant errors relative to the fiducial single-lens case.
Again, the statistical errors differ by less than a factor of $2$ from the
fiducial estimate in equation~(\ref{eqn:scale}).

\subsection{Finite Unresolved Flux\label{sec:bg}}

\begin{figure*}[t]   
\epsscale{1.0} \centerline{\plotone{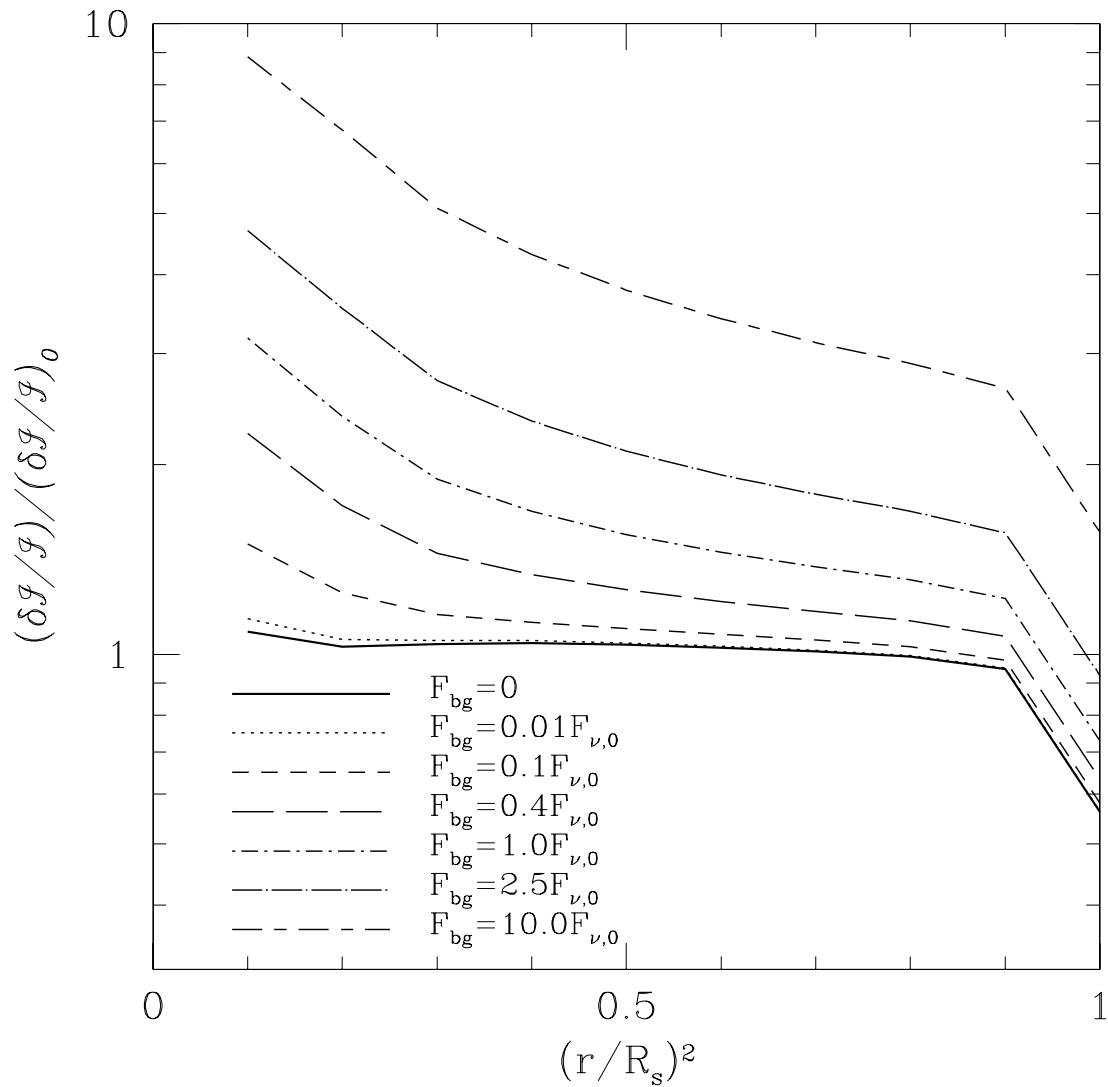}}
\caption{ \footnotesize The fractional error in the recovered
intensity profile $\delta\ci/\ci$ normalized to the
fiducial error $(\delta\ci/\ci)_0$ given by the scaling relation in
\eq{eqn:scale}, as a function of the square of the normalized radius
of the afterglow for various assumptions about the magnitude the
unresolved background flux $F_{\rm bg}$ relative to the flux
$F_{\nu,0}$ of the afterglow at $t=1~\day$.}

\label{fig:fig10}
\end{figure*}

Observed afterglows often show evidence for a flattening of the light
curve to a constant flux at late times.  This is usually interpreted
as due to unresolved flux from the host galaxy of the GRB.  Even if
the flux from the host galaxy is small, the sky background will
eventually dominate over that of the afterglow. In other words, it is
inevitable that $F_{\nu}(t) < F_{\rm bg}$ at late times.  For ground-based 
infrared observations, the background flux is likely to be larger than the
afterglow flux at all relevant times.  
A finite background flux will
increase the errors on the recovered parameters $\ci_i$, and
preferentially so for smaller radii, since the information on smaller
radii comes at later times, when the background is more significant.
We explore this effect by introducing a background flux which amounts
to some fraction of $F_{\nu,0}$ for the afterglow.  \fig{fig:fig10}
shows the expected fractional errors on the recovered intensity
profile relative to the fiducial value in equation~(\ref{eqn:scale}),
assuming a power-law flux index of $\alpha=1$ for the afterglow, for
relative background fluxes of $F_{\rm bg}/F_{\nu,0}= 0.01, 0.1, 0.4,
1.0, 2.5$, and $10.0$. At optical-infrared frequencies, this
corresponds to a background of magnitude $5, 2.5, 1, 0, -1, -2.5$ 
relative to the (unlensed) afterglow at $1~\day$.  
As expected, the errors generally increase as the background flux
increases, especially for smaller radii.  However, unless the background
flux is higher than $F_{\nu,0}$, the expected errors are $\lsim 3$ times
greater than the fiducial estimate.  
In other words, for optical frequencies, 
the background flux should not compromise the measurement of the intensity profile.  However, for, e.g., ground-based observations in the near-infrared, where the sky background can be many magnitudes brighter than the afterglow flux at 1 day, the accuracy with which the intensity profile can be measured will considerably larger than the fiducial estimate (\Eq{eqn:scale}). Thus spaced-based near-infrared observations would improve the accuracy of the recovered intensity profile.

\subsection{A Worked Example\label{sec:discussion}}

\begin{figure*}[t]   
\epsscale{1.0} \centerline{\plotone{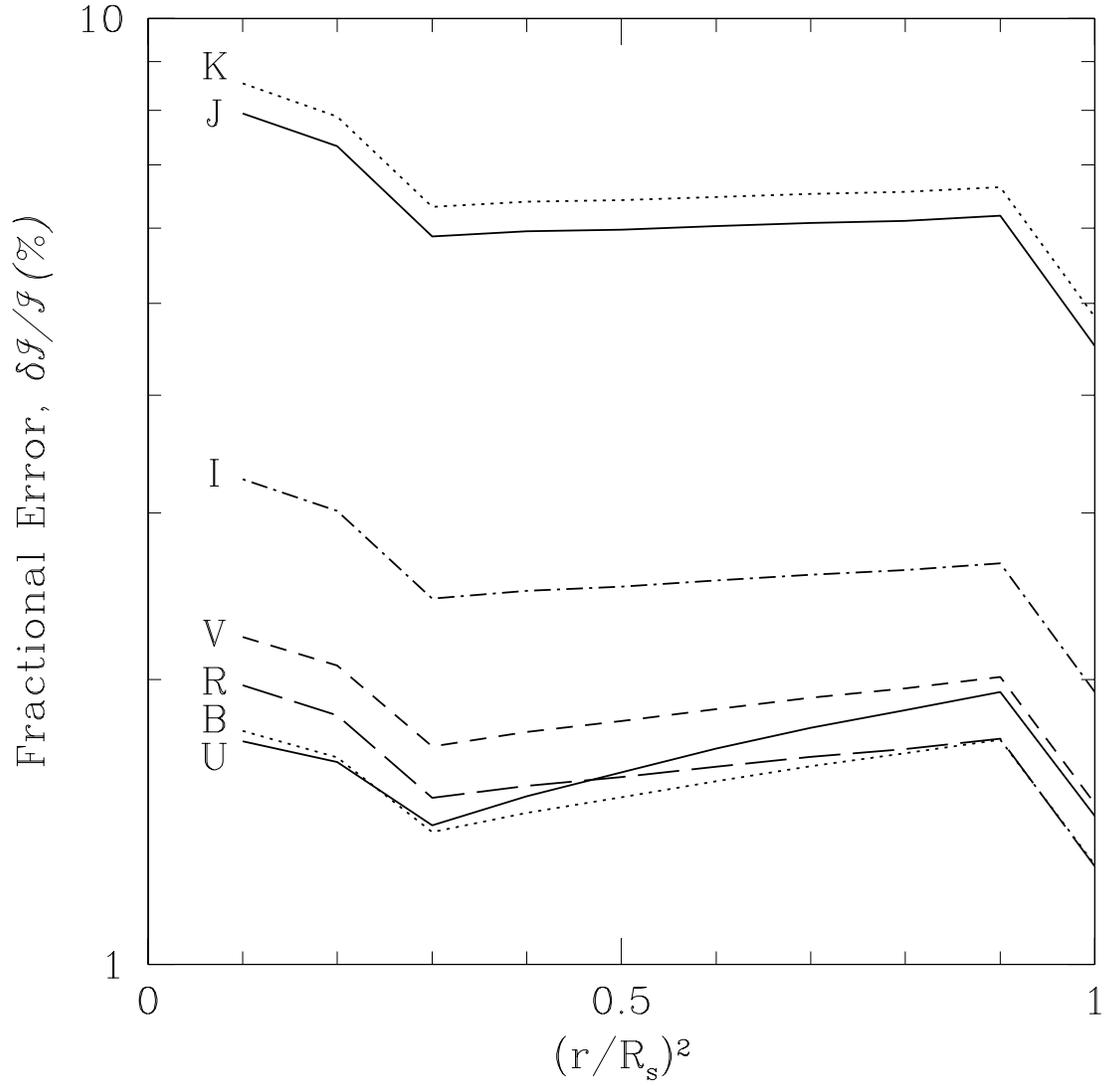}}
\caption{ \footnotesize The fractional error $\delta\ci/\ci$ in the
recovered relative intensity profile in percent as a function of the square
of the normalized radius $X=r/R_{\rm s}$, for the afterglow and
microlensing parameters of GRB 000301C, assuming a uniform source, and that
observations were made with a $4~{\rm m}$ telescope from the peak of the
microlensing event until $\sim 30$ days after the peak.  }
\label{fig:fig12} 
\end{figure*}

As discussed in \S\ \ref{sec:scaling}, the results we have presented so far
are quite general.  We now estimate the errors expected under realistic
observational conditions and a realistic model.  To do so, we make
estimates of the various terms in \eq{eqn:scale} by adopting the parameters
for the observed afterglow of GRB 000301C.  \citet{gls2000} fit the optical
and infrared photometry of GRB 000301C to double power-law flux decline
magnified by the simple microlensing model of \citet{landp1998}.  They find
best-fit parameters of $\alpha_1=1.1$ and $\alpha_2=2.9$ for the power-law
flux indices, similar to the values found originally by \citet{sagar2000}
without including the microlensing modification.  For simplicity, we ignore
the break, and simply adopt a single power-law of slope $\alpha=1.1$.  We
adopt the best-fit microlensing parameters for the entire dataset of
$R_0=0.49$ and $b=1.04$\footnote{\citet{ggl2001} find the impact parameter
in GRB 000301C is not well-constrained, but is in the range $0.2 \lsim
b\lsim 0.7$, somewhat smaller than that found by \citet{gls2000}.  For
illustrative purposes, we will simply adopt the \citet{gls2000} value of
$b=1.04$.}.  Finally, we extrapolate by eye the unlensed flux of the
afterglow back to $t=1~{\rm day}$ for the $U,B,V,R,I,J$ and $K$-bands.
Table 1 shows the demagnified, extrapolated magnitudes at $t=1~\day$ of the
afterglow of GRB 000301C, along with the assumed sky brightness, and
$\Gamma_\nu/A_{\rm T}$, the number of photons per second per square meter
of collecting telescope area $A_{\rm T}$.  Note that, for $R_0=0.49$ and
$b=1.04$, the peak of the microlensing event (for a uniform source) occurs
at $\tp=(b/R_0)^{8/5} \approx 3~\days$.  We calculate the expected errors
$\delta\ci/\ci$ assuming a 4-m telescope, a seeing of $1^{\prime\prime}$, a
uniform source, and observations between the peak, $t=\tp=3~\days$ and
$t\simeq 9\tp\simeq 30~\days$.  We adopt a uniform source in order to
provide a model-independent estimate of the fractional errors expected for
realistic observing conditions.  Image profiles in the optical and infrared
are unlikely to be well-represented by a uniform source \citep{gl2001},
however this will not affect the errors in the recovered profile by more
than a factor of $\sim 3$ (see \S \ref{sec:iprofs}).  The results are shown
in \fig{fig:fig12}.  For observations in the optical, the relative
intensity profile can be recovered accurately, $\delta\ci/\ci \lsim 3\%$.
For observations in the infrared, where the sky background is dominant, the
expected errors are $\sim 10\%$.  Although we do not specifically consider
other wavelengths here, we note that radio afterglows peak later and last
longer than afterglows at shorter wavelengths (\citealt{kulk2000} and
references therein).  This, combined with the lower energy of the photons,
implies that the photon flux at a given time is larger in the radio than in
the optical.  At sufficiently high radio frequencies where the effect of
scintillations is sub-dominant, radio observations may provide more
accurate measurements of the afterglow image.  Since the afterglow image is
expected to be different at radio frequencies \citep{gl2001}, we advocate
performing measurements over as wide a range of frequencies as possible.

\begin{figure*}[t]   
\epsscale{1.0} \centerline{\plotone{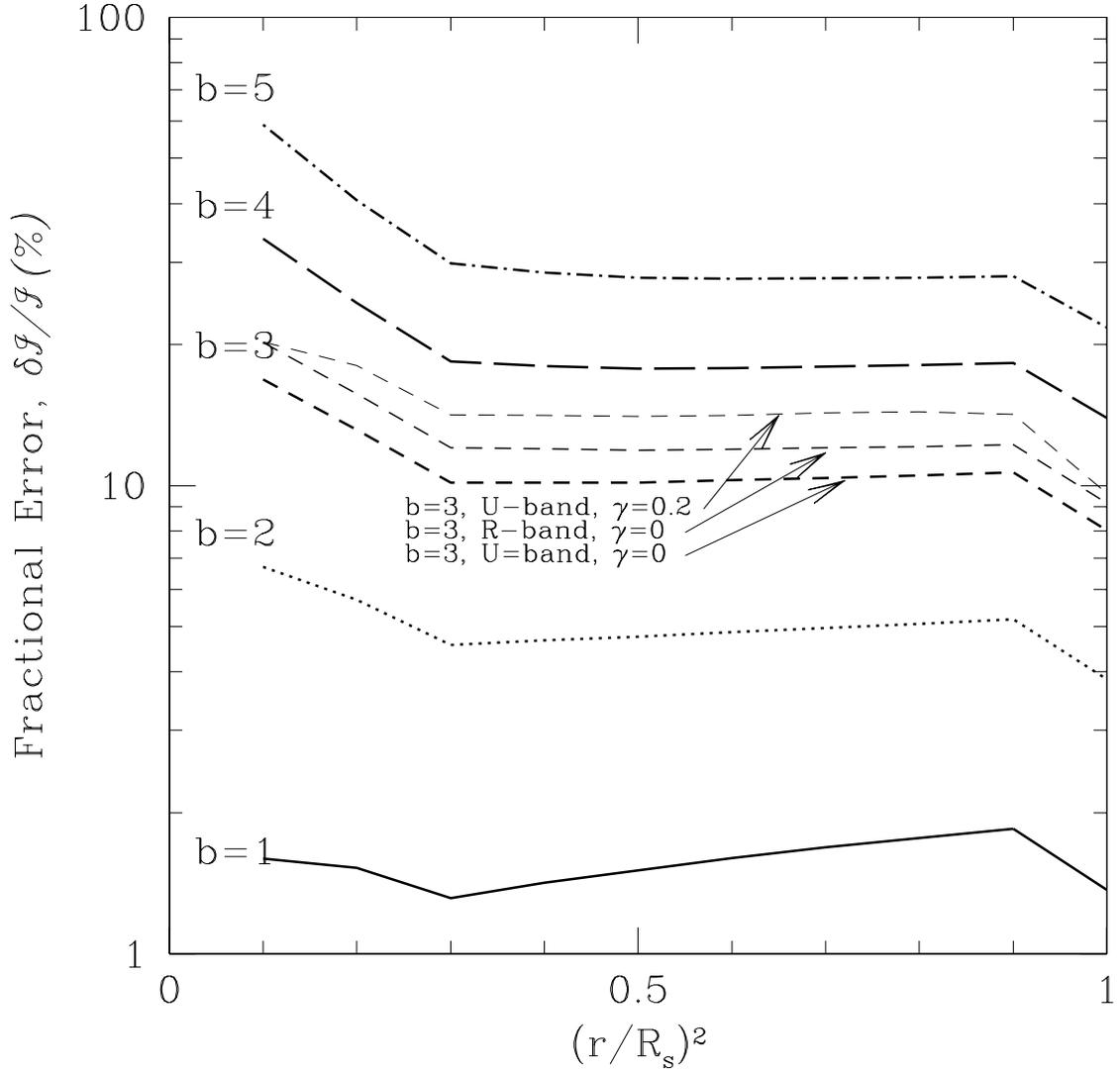}}
\caption{ \footnotesize Same as Figure~\ref{fig:fig12} for $U$-band,
except we have varied the impact parameter of the microlensing event,
$b$.  Note that the fraction of all afterglows with impact parameter
$\le b$ is $\sim 1\% b^2$.}
\label{fig:fig13} 
\end{figure*}

Given that the fraction of afterglows that exhibit microlensing events with
impact parameter $\le b$ scales as $b^2$, it is interesting to consider
what kind of errors one would expect for the specific example of GRB
000301C considered above, but with larger impact parameters.  We therefore
repeat the calculation above with the same parameters, except we now vary
the impact parameter.  Specifically, we consider $b=1,2,3,4$, and $5$, as
would be expected in $\sim 1\%, 4\%, 9\%, 16\%$, and $25\%$ of all GRB
afterglows.  In \fig{fig:fig13} we show the results for the $U$-band (where
the background is the smallest), assuming that observations are taken from
peak, $t=\tp=3 b^{5/8}~\days$ until $t\simeq 9\tp\simeq 30 b^{5/8}~\days$.

For $b\le 3$, the fractional errors in the recovered intensity profile are
$\lsim 10\%$ for the majority of the area of the image.  Thus, it should be
possible to recover the intensity profile to an accuracy of $\lsim 20\%$
for $\sim 10\%$ of all afterglows.  To test the robustness of this
conclusion, we also show in\fig{fig:fig13} the results for the $b=3$
case, but considering a band with larger background flux ($R$-band), and
also a non-zero external shear ($\gamma=0.2$ with source on the shear
axis).  We find that the error is not greatly inflated, $\delta\ci/\ci
\lsim 20\%$.  Of course, these results assume that the afterglow can be
monitored for many months after the peak of the microlensing event.
Practical aspects aside, such monitoring may not be possible even in
principle in those cases where the emission is due to a highly collimated
jet.  Such afterglows will eventually exhibit breaks in the light curve and
changes in the image structure when the Lorentz factor of the fireball
falls below the inverse of the opening angle of the jet.  In this case, the
errors on the annuli centered at smaller radii would be inflated (see \S
\ref{sec:dobs}).  Additional deviations from our results may occur as the
fireball decelerates to non-relativistic speeds \citep{mandl2001}.

\section{Conclusions\label{sec:conclusions}}

Microlensing offers a unique method for probing the dynamics of GRB
fireballs.  As the GRB afterglow image expands, it is differentially
magnified by the lens. Different annuli of the image are resolved as they
cross the lens position at different times.  Thus the light curve of the
afterglow during the microlensing event can be inverted to obtain the
radial intensity profile $\ci(X)$, where $X=r/R_{\rm s}$ is the fractional
radius of the circular image.  We developed the formalism necessary to
calculate the expected fractional errors on the recovered intensity profile
$\delta\ci/\ci$.

Assuming continuous observations, uniform sources, single lenses, and
zero background flux, we find that the expected errors follow a
general scaling relation (Eq.~\ref{eqn:scale}) in terms of the
afterglow flux, afterglow radius in units of the Einstein radius, the
angular separation between the afterglow and the lens, and the number
of resolution elements in the recovered intensity profile.
We have tested the accuracy of this relation by, in turn, relaxing the
assumptions of continuous observations, uniform sources, single
lenses, and zero background flux.  We find that, for reasonable
choices of the relevant parameters, none of these refinements changes
the expected errors by more than a factor of $\sim 3$.  Therefore, the
fiducial scaling relation is fairly robust.

Notably, we have found that observations starting at the peak $\tp$ of the
microlensing event and ending at $7\tp$ result in almost the same accuracy
as measurements of the entire afterglow.  Therefore, one can monitor
frequently {\it only} those afterglows that are significantly microlensed
by ``alerting'' to these events before they reach their peak.

Finally, we calculated the errors expected for observations in the
near-infrared and optical regimes of a specific example.  Adopting the
observed and inferred parameters for the (possibly) microlensed
optical/infrared afterglow of GRB 000301C, we find that the relative
intensity profile can typically be measured with a resolution of 10 and
accuracy of ${\cal O}(1\%)$ in the optical and ${\cal O}(10\%)$ in the
infrared using 4m-class telescopes.  Fluctuations in the light curve due to
inhomogeneities in the ambient gas distribution are typically small and
should not compromise this precision \citep{wl2000,hal2000}.

In the course of this study, we have had to make a number of simplifying
assumptions.  These include: (i) the unlensed flux of the afterglow is a
power-law function of time, (ii) the radius of the afterglow image scales
as a power-law of time $R_{\rm s}\propto t^{\delta}$, (iii) the afterglow
image is circularly symmetric, (iv) the intensity profile is self--similar,
i.e. $\ci(X)$ is independent of time, (v) the lens resides in a region of
relatively low optical depth, and (vi) the covariances between the surface
brightness profile parameters and the afterglow and microlensing parameters
are small.  The first assumption is a generic prediction of the simplest
afterglow models, provided that the observed frequency does not cross one
of the several spectral breaks, and that $1/\Gamma$ is smaller than the
collimation angle of the ejecta.  When a break in the light curve occurs
--- either due to the crossing of a spectral break or jet effects --- the
intensity profile is also likely to change, thus also violating assumption
(iv).  When the break is due to jet effects, or when the fireball becomes
non-relativistic, then the index $\delta$ of the power-law form for the
image radius will change.  Furthermore, breaks in the light curve will
produce additional uncertainties in the microlensing and afterglow
parameters by introducing additional correlations.  For example, in the
case of a single power-law flux decline, the impact parameter $b$ can be
found by extrapolating the late-time light curve (when the afterglow is not
lensed) to very early times, when the afterglow is lensed by an amount that
depends only on $b$.  Clearly, this procedure becomes more uncertain when
the light curve exhibits a break.  The assumption of circular symmetry may
be crude at best, especially for collimated outflows.  For single lenses,
it is clear that the information gathered about the surface brightness
profile is inherently one--dimensional, and therefore departures from
circular symmetry will be difficult to detect.  Therefore, the surface
brightness profile inferred assuming circular symmetry will be biased. In
principle, binary lenses or single lenses with external shear would provide
more information about departures from circular symmetry than a simple
single lens; in practice, however, the additional freedom introduce by the
more complex lens models may make extracting this information difficult.
Our assumption of small optical depth is likely to hold in the outskirts of
galaxies.  For sight--lines passing near the center of an intervening
galaxy, the magnification structure of the lens is unlikely to be
well--represented by any of the three lenses we considered: an isolated
single lens, a single lens with external shear, or a binary--lens.  In the
high-optical depth regime, it may be difficult to reconstruct the relative
intensity profile directly, due to the complexity of the magnification
structure.  Finally, the assumption that the covariances between the
intensity profile parameters and the afterglow and microlensing parameters
are small is unlikely to hold in practice.  This means that the errors we
calculate are effectively lower limits to the true errors.  The effects of
these various assumptions on the resultant errors on the recovered
intensity profile are interesting and important topics for future
study. 

Our basic results are encouraging.  Errors of $1\%$ would likely provide
very stringent constraints on afterglow models, and even errors of $\sim
10\%$ would be interesting (see Fig. 2 in Granot \& Loeb 2001).  Errors of
the latter magnitude can be obtained from events with larger impact
parameters.  For the specific afterglow parameters adopted above, we find
errors of $\delta\ci/\ci\sim 6\%, 10\%, 20\%$ in the $U$-band for impact
parameters in units of $\thetae$ of $b=2,3$ and $4$.  Since the number of
expected events scales as $b^2$, this implies that interesting results
might be obtained for a significant sample of afterglows.  The forthcoming
Swift satellite\footnote{Planned for launch in 2003; see
http://swift.sonoma.edu/} could provide hundreds of afterglow targets per
year.  A network of 1m class telescopes similar to that used in
gravitational microlensing searches of the Local Group
\citep{albrow1998,mps1999} would provide an ideal method for ``alerting''
the community to {\it microlensed} afterglows, which could in turn be
intensively monitored to provide information about the structure of the
afterglow image.  Although this represents a considerable expenditure of
resources, the information gained would be invaluable, and furthermore
cannot be currently acquired by any other method.

\section*{Acknowledgements}
We thank Andrei Gruzinov for helpful discussions. This work was supported
in part (for SG) by NASA through a Hubble Fellowship grant from the Space
Telescope Science Institute, which is operated by the Association of
Universities for Research in Astronomy, Inc., under NASA contract
NAS5-26555, and (for AL) by grants from the Israel-US BSF (BSF-9800343),
NSF (AST-9900877), and NASA (NAG5-7039).

\begin{table*}
\begin{center}
\begin{tabular}{|c|c|c|c|}
\tableline Filter & Apparent & $\mu_{\rm sky}$ & $\Gamma_\nu/A_{\rm T}
$\tablenotemark{2} \\ & Magnitude\tablenotemark{1} & $({\rm mag~asec^{-2}})$ &
$(\gamma~{\rm s^{-1} m^{-2}})$ \\ \tableline \tableline $U$ & 20.1 &
22.8 &43 \\ $B$ & 20.6 & 22.5 &77 \\ $V$ & 20.1 & 21.5 &78 \\ $R$ &
19.7 & 20.8 &130 \\ $I$ & 19.3 & 19.3 &128 \\ $J$ & 18.3 & 15.9 &194
\\ $K$ & 16.8 & 13.7 &540 \\ \tableline
\end{tabular}
\end{center}
\tablenum{1} {\bf Table 1} Adopted fluxes for GRB 000301C.
\tablenotetext{1} {Demagnified apparent magnitude of GRB 000301C
extrapolated to 1 day after the burst.}  \tablenotetext{2} {Number of
photons per second per square meter of collecting area $A_{\rm T}$.}
\label{tbl:table1}
\end{table*}

\clearpage

\clearpage

\clearpage

\clearpage

\clearpage

\clearpage

\clearpage

\clearpage

\clearpage

\begin{thebibliography}{}

\bibitem[Afonso et~al.(2000)]{afonso2000} Afonso, C., et al.\ 2000,
ApJ, 532, 340

\bibitem[Albrow et~al.(1998)]{albrow1998} Albrow, M., et al.\ 1998,
ApJ, 509, 687

\bibitem[Albrow et~al.(1999a)]{albrow1999} Albrow, M., et al.\ 1999a,
ApJ, 522, 1011

\bibitem[Albrow et~al.(1999b)]{albrow1999b} Albrow, M., et al.\ 1999b,
ApJ, 522, 1022

\bibitem[Albrow et~al.(2000)]{albrow2000} Albrow, M., et al.\ 2000,
ApJ, 534, 894

\bibitem[Albrow et~al.(2001a)]{albrow2001a} Albrow, M., et al.\ 2001a,
ApJ, 549, 000

\bibitem[Albrow et~al.(2001b)]{albrow2001b} Albrow, M., et al.\ 2001b,
ApJ, 550, L173

\bibitem[Alcock et~al.(1996)]{alcock1996} Alcock, C., et al.\ 1996,
ApJ, 463, L67

\bibitem[Alcock et~al.(1997)]{alcock1997} Alcock, C., et al.\ 1997,
ApJ, 491, 436

\bibitem[Alcock et~al.(2000)]{alcock2000} Alcock, C. et al. 2000, ApJ,
542, 281

\bibitem[Blaes \& Webster(1992)]{bandw1992} Blaes, O.M., \& Webster,
R.L.\ 1992, ApJ, 391, L63

\bibitem[Blandford \& McKee(1976)]{bm1976} Blandford, R. D., \& McKee,
C. F. 1976, Phys. Fluids, 19, 1130

\bibitem[Castro et~al.(2001)]{castro2001} Castro, S.M., Pogge, R.W.,
Rich, R.M., DePoy, D.L., \& Gould, A.\ 2001, ApJ, 548, L197

\bibitem[Chang \& Refsdal(1979)]{cha79} Chang, K., \& Refsdal, S.\
1979, Nature, 282, 561

\bibitem[di Stefano \& Perna(1997)]{dip1997} di Stefano, R., \& Perna,
R.\ 1997, ApJ, 488, 55

\bibitem[Fruchter et~al.(1999)]{fruchter1999} Fruchter, A.S., et~al.\
1999, ApJ, 516, 683

\bibitem[Frail et~al.(1997)]{frail1997} Frail, D.A., et~al.\ 1997,
Nature, 389, 261

\bibitem[Freedman \& Waxman(1999)]{fandw1999} Freedman, D. L., \&
Waxman, E. 1999, ApJ, submitted (astro-ph/9912214)

\bibitem[Fukugita, Hogan, \& Peebles(1998)]{fhp1998} Fukugita, M.,
Hogan, C. J., \& Peebles, P. J. E. 1998, ApJ, 503, 518

\bibitem[Gaudi \& Gould(1999)]{gandg1999} Gaudi, B.S., \& Gould, A.\
1999, ApJ, 513, 619

\bibitem[Gaudi, Granot, \& Loeb(2001)]{ggl2001}
Gaudi, B.S., Granot, J., \& Loeb, A.\ 2001, ApJL, in preparation

\bibitem[Garnavich, Loeb \& Stanek(2000)]{gls2000} Garnavich, P.M.,
Loeb, A., \& Stanek, K.Z.\ 2000, ApJ, 544, L11

\bibitem[Goodman(1997)]{g1997} Goodman, J. 1997, New Astronomy, 2, 449

\bibitem[Gould(2001)]{gould2001} Gould, A.\ 2001, PASP, in press (astro-ph/0103516)

\bibitem[Granot \& Loeb(2001)]{gl2001} 
Granot, J., \& Loeb, A. 2001, ApJ, 551, L63

\bibitem[Granot, Piran \& Sari(1999a)]{gps1999} Granot, J., Piran, T.,
\& Sari, R.\ 1999a, ApJ, 513, 679

\bibitem[Granot, Piran \& Sari(1999b)]{gps1999b}
-------------------------------.\ 1999b, ApJ, 527, 236

\bibitem[Halpern et~al.(2000)]{hal2000} Halpern, J., et al.\ 2000,
ApJ, 543, 697

\bibitem[Heyrovsk{\' y}(2001)]{hey2001} Heyrovsk{\' y}, D.\ 2001, ApJ,
submitted

\bibitem[Heyrovsk{\' y}, Sasselov, \& Loeb(2000)]{hsl2000} Heyrovsk{\'
y}, D., Sasselov, D., \& Loeb, A.\ 2000, ApJ, 543, 406

\bibitem[Katz \& Piran(1997)]{kandp1997} Katz, J., \& Piran, T.\ 1997,
ApJ, 490, 772

\bibitem[Koopmans \& Wambsganss(2001)]{kandw2001} Koopmans, L.V.E., \&
Wambsganss, J.\ 2001, MNRAS, in press (astro-ph/0011029)

\bibitem[Kulkarni et~al.(2000)]{kulk2000} Kulkarni, S.R., et~al.\
2000, in AIP Conf.\ Proc., Gamma-Ray Bursts: 5th Huntsville Symposium,
ed. R. M. Kippen, R. S. Mallozzi, \& G. J. Fishman (New York: AIP)

\bibitem[Loeb \& Perna(1998)]{landp1998} Loeb, A., \& Perna, R.\ 1998,
ApJ, 495, 597

\bibitem[Loeb \& Sasselov(1995)]{ls1995} 
Loeb, A., \& Sasselov, D.  1995, ApJ, 449, L33

\bibitem[Mao \& di Stefano(1995)]{mdi1995} Mao, S., \& di Stefano, R.\
1995, ApJ, 440, 22

\bibitem[Mao \& Loeb(2001)]{mandl2001} Mao, S., \& Loeb, A.\ 2001,
ApJ, 547, L97

\bibitem[Meszaros \& Rees(1997)]{mr1997} Meszaros, P., \& Rees,
M. J. 1997, ApJ, 476, 232

\bibitem[Paczynski \& Rhoads(1993)]{pr1993} Paczynski, B., \& Rhoads,
J. E. 1993, ApJ, 418, L5

\bibitem[Panaitescu(2001)]{pana2001}
Panaitescu, A.\ 2001, ApJ, in press (astro-ph/0102401)

\bibitem[Panaitescu \& Kumar(2001)]{pandk2001} Panaitescu, A., \&
Kumar, P.\ 2001, ApJ, in press (astro-ph/0010257)

\bibitem[Panaitescu \& Meszaros(1998)]{pm1998} Panaitescu, A., \&
Meszaros, P.\ 1998, 493, L31

\bibitem[Pian et~al.(1998)]{pian1998} Pian, E., et~al.\ 1998, ApJ,
492, L103

\bibitem[Press \& Gunn(1973)]{pandg1973} Press, W.H., \& Gunn, J.E.\
1973, ApJ, 185, 397

\bibitem[Rhie et~al.(1999)]{mps1999} Rhie, S.H., Becker, A.C.,
Bennett, D.P., Fragile, P.C., Johnson, B.R., King, L.J., Peterson,
B.A., \& Quinn, J.\ 1999, ApJ, 522, 1037

\bibitem[Rhoads(1997)]{rhoads1997} Rhoads, J. E. 1997, ApJ, 487, L1

\bibitem[Rhoads \& Fruchter(2000)]{randf2000} Rhoads, J., \& Fruchter,
A.S.\ 2000, ApJ, in press

\bibitem[Sagar et~al.(2000)]{sagar2000} Sagar, R., Mohan, V., Pandey,
S.B., Pandey, A.K., Stalin, C.S., \& Tirado, A.J.\ 2000, BASI, 28, 499

\bibitem[Sasselov(1997)]{sass1997} Sasselov, D.\ 1997, in Variable
Stars and the Astrophysical Returns of Microlensing Surveys, ed.\
R. Ferlet, J.-P. Maillard, \& B. Raban (Gif-sur-Yvette: Editions
Fronti{\` e}res), 141

\bibitem[Sari (1998)]{sari1998} Sari, R. 1998, ApJ, 494, L49

\bibitem[Sari, Piran, \& Narayan(1999)]{spn1999} Sari, R., Piran, T.,
\& Narayan, R. 1998, ApJ, 497, L17

\bibitem[Schneider \& Weiss(1986)]{sandw1986} Schneider, P., \& Weiss,
A.\ 1986, 164, 237

\bibitem[Schneider, Ehlers \& Falco(1992)]{bible} Schneider, P.,
Ehlers, J., \& Falco, E.E.\ 1992, Gravitational Lenses (Heidelberg:
Springer), p. 313

\bibitem[Udalski et~al.(1994)]{udal1994} Udalski, A., Szyma{\' n}ski,
M., Kalu{\. z}ny, J., Kubiak, M., Mateo, M., Krzemi{\' n}ski, W., \&
Paczy{\' n}ski, B.\ 1994, AcA, 44, 227

\bibitem[Valls-Gabaud(1995)]{valls1995} Valls-Gabaud, D.\ 1995, in
Large Scale Structure of the Universe, ed.\ J.P. M{\" u}cket,
S. Gottl{\" o}ber, \& V. M{\" u}ller (Singapore: World Scientific),
326

\bibitem[Wang \& Loeb(2000)]{wl2000} Wang, X., \& Loeb, A. 2000, ApJ,
535, 788

\bibitem[Wijers \& Galama(1999)]{wg1999} Wijers, R. A. M. J., \&
Galama, T. J. 1999, ApJ, 523, 177

\bibitem[Witt(1990)]{witt1990} Witt, H.\ 1990, A\&A, 236, 311

\bibitem[Witt \& Mao(1994)]{wandm1994} Witt, H.J., \& Mao, S.\ 1994,
ApJ, 430, 505

\bibitem[Wambsganss(1997)]{wambs1997} Wambsganss, J.\ 1997, MNRAS,
284, 172

\bibitem[Waxman(1997a)]{wax1997a} Waxman, E.\ 1997a, ApJ, 485, L5

\bibitem[Waxman(1997b)]{wax1997b} Waxman, E.\ 1997b, ApJ, 489, L33

\bibitem[Waxman(1997c)]{wax1997c} Waxman, E.\ 1997c, ApJ, 491, L19

\bibitem[Waxman, Kulkarni, \& Frail(1998)]{wkf1998} Waxman, E.,
Kulkarni, S. R., \& Frail, D. A. 1998, ApJ, 497, 288

\end{thebibliography}
\end{document}